\documentstyle[aaspp4,epsfig,flushrt]{article}

\newcommand{\dpp}{\frac{\delta p}{p}}
\newcommand{\gamu}{\frac{1}{\Gamma_1}}
\newcommand{\dlnz}[1]{\frac{d\ln #1}{dz}}
\newcommand{\gcm}{\,g\,cm^{-3}}
\newcommand{\keV}{\,{\rm keV}}
\newcommand{\secd}{\,{\rm s}}
\newcommand{\cm}{\,{\rm cm}}
\newcommand{\Hz}{\,{\rm Hz}}
\newcommand{\EF}{E_F}

\begin{document}

\title{Hydrogen Electron Capture in Accreting Neutron Stars and the
Resulting g-Mode Oscillation Spectrum}

\author{Lars Bildsten and Andrew Cumming} 
\affil{Department of Physics and Department of Astronomy\\ University
of California, Berkeley, CA 94720\\
bildsten@fire.berkeley.edu, cumming@fire.berkeley.edu}

\begin{abstract}

  We investigate hydrogen electron capture in the ocean of neutron
stars accreting at rates 
$10^{-10}\lesssim\dot M\lesssim 10^{-8} M_\odot \ {\rm yr^{-1}}$.
These stars burn the accreted hydrogen and helium unstably
in the upper atmosphere ($\rho \lesssim \ 10^6 \ {\rm g \ cm^{-3}}$)
and accumulate material which usually contains some small amounts of
hydrogen (mass fractions are typically $X_r\approx 0.1-0.2$) mixed in
with the heavier iron group ashes. The subsequent evolution of this
matter is determined by compression towards higher densities until
electron capture on the hydrogen occurs. We construct steady-state
models of the electron captures and the subsequent neutron
recombinations onto the heavy nuclei. The density discontinuity from
these captures gives rise to a new g-mode (much like a surface wave),
which has a lowest order ($l=1$) frequency of $\approx 35 \ {\rm
Hz}(X_r/0.1)^{1/2}$ on a slowly rotating ($f_s\ll 30 \ {\rm Hz}$)
star. We also discuss, for the first time, a new set of non-radial
g-modes unique to these high accretion rate neutron stars. These modes
are ``trapped'' in the finite thickness layer where the electron
captures are occurring. The lowest order mode frequencies are in the
$1-10 \ {\rm Hz}$ range for a few radial nodes on a slowly rotating
star. Though the majority of the mode energy resides in the electron
capture transition layer, the eigenmode propagates to higher altitudes
above the layer and can thus be potentially observable and excited by
the nuclear burning or other mechanisms. We also show that the density
jump splits the ocean's thermal modes into two distinct sets, which
have most of their energy either above or below the discontinuity.
We conclude by discussing how the dispersion relations for these modes
are modified for a rapidly rotating ($f_s\gg 30 \ {\rm Hz}$) neutron
star. Whether any of these modes are observable depends on the damping
mechanism and the ability to excite them, issues we will address in a
future paper.

\end{abstract}


\begin{center}
\bf To appear in The Astrophysical Journal, October 20, 1998, Vol. 506
\end{center}

\vfill\eject 

\section{Introduction}
  
 The liquid core, solid crust and overlying ocean and atmosphere give
a neutron star (NS) a very rich spectrum of non-radial oscillations,
with acoustic p-modes in the 10 kHz range and a variety of lower
frequency g-modes from the toroidal and spheroidal displacements of
the crust and surface layers (see McDermott, Van Horn
\& Hansen 1988 for an overview). Non-radial g-mode oscillations have
been studied extensively for isolated and slowly rotating radio
pulsars, where the atmosphere and ocean are quiescent. These studies
can be separated by the sources of buoyancy that were considered: (1)
entropy gradients (McDermott, Van Horn \& Scholl 1983; McDermott et
al. 1988), (2) density discontinuities (Finn 1987; McDermott 1990;
Strohmayer 1993) and (3) mean molecular weight gradients due to
$\beta$--equilibrium in the core (Reisenegger \& Goldreich
1992). Unfortunately, none of these modes have been securely
identified with any particular radio pulsar phenomenon.

There are also many accreting neutron stars in binary systems that
emit much of their accretion luminosity in X-rays (Lewin, van Paradijs
\& van den Heuvel 1995). The thermal and
compositional structure of these accreting NSs is quite different from
that of the isolated radio pulsars, and is far from equilibrium.
There have been previous studies of non-radial oscillations in these
objects (McDermott \& Taam 1987; Bildsten \& Cutler 1995 (hereafter
BC95); Bildsten, Ushomirsky \& Cutler 1996 (hereafter BUC96);
Strohmayer \& Lee 1996 (hereafter (SL96)).  In this paper, we
calculate the g-mode spectrum of the ocean of an accreting NS,
focusing on the critical effects of hydrogen electron capture.

\subsection{ Observational Motivation} 

Much of this work is motivated by the rich spectrum of quasi-periodic
oscillations (QPOs) seen in the luminosity of the brightest accreting
neutron stars. Utilizing {\it EXOSAT} data, Hasinger \& van der Klis
(1989) found that the highest accretion rate objects ($\dot M>
10^{-10} M_\odot \ {\rm yr^{-1}}$) fall into two separate classes. Six
objects trace out all or part of a ``Z'' in an X-ray color-color
diagram and exhibit time-dependent behavior that correlates with the
position along the Z. These ``Z sources'' have QPO's in the 15--50 Hz
and 5--7 Hz range with up to 10\% modulation of the X-ray flux. The
other objects fall into separated regions of the color-color diagram
and do not show this distinctive QPO phenomenology.  These ``Atoll
sources'' accrete at lower rates than the Z sources and exhibit Type I
X-ray bursts resulting from the unstable ignition of the accumulated
hydrogen and helium (see Bildsten 1998a for a recent review).

The {\it Rossi X-Ray Timing Explorer} ({\it RXTE}) has recently added
to this list of QPO phenomenology. Two drifting kHz QPOs are seen in
the persistent emission from both Atoll and Z sources, and coherent
oscillations have been observed during Type I X-ray bursts from
several Atoll sources (see Strohmayer et al. 1996 for an example).  In
these objects, the kHz QPOs are separated by a frequency identical to
that seen in the burst, leading naturally to a beat frequency model in
which the difference frequency is the NS spin frequency (Strohmayer et
al. 1996). The physical origin of the upper frequency differs in the
models (see van der Klis 1998 for a summary), and usually involves the
Kepler frequency at some special place in the accretion disk (Kaaret,
Ford \& Chen 1997; Zhang, Strohmayer \& Swank 1997; Miller, Lamb \&
Psaltis 1998). The inferred spin frequencies are in a remarkably
narrow range, 250--500 Hz (White \& Zhang 1997; van der Klis 1998;
Bildsten 1998b).  There are also many QPOs with frequencies less than
100 Hz seen in the Atoll sources. Stella \& Vietri (1998) proposed
that these might be due to Lense-Thirring precession of the inner
accretion disk.

It may be that all of the $<100$ Hz and kHz QPOs can be accommodated
within the existing models (see van der Klis 1995, 1998) that use the
accretion disk and/or spherical flow to generate the periodic
phenomena. However, it is important to pursue non-radial oscillations
of the surface layers of the NS as a possible source of some of these
periodicities. The predicted frequencies are well-constrained,
depend on the underlying NS structure and have well-understood
dispersion relations, even when the NS is rapidly rotating
(BUC96). This means that the successful identification of a non-radial
pulsation with an observed frequency would tell us much about the
thermal and compositional makeup of the NS and measure or constrain
the rotation rate and magnetic field.

\subsection{Accreting Neutron Star Structure and Non-Radial Oscillations}

In most cases where stellar pulsations are studied, the underlying
stellar model is reasonably well understood and constrained via other
observations. Initial theoretical work then consists of finding the
adiabatic mode structure and exploring possible excitation mechanisms,
usually focusing on the $\kappa$-mechanism which relies on opacity
changes during the pulsations (see Cox 1980 for an excellent
overview). The situation is quite different in an accreting NS. The
conditions in the outer layers can change on short timescales, the
deep internal structure depends on the accretion history and there is
a broader range of excitation possibilities.  We now briefly outline
the underlying NS structure and the resulting g-mode spectrum.

\begin{figure}[hbp]
\centering{\epsfig{file=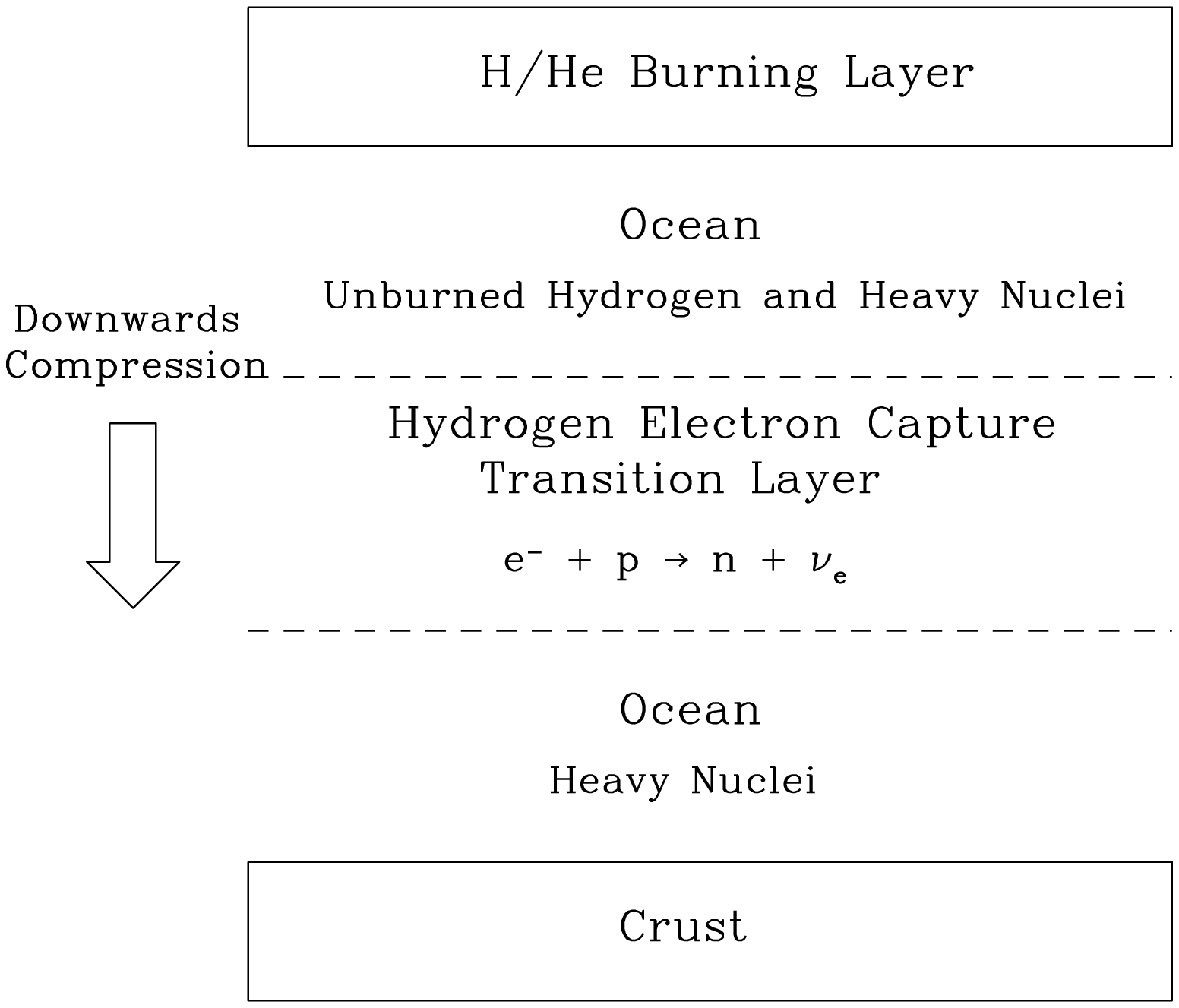,width=3.5 in}}
\caption{
A schematic of the atmosphere, ocean and crust of a neutron star
accreting at a rapid rate. The accreted H/He burns at a column depth
$y\approx 10^8\ {\rm g\ cm^{-2}}$ or a density $\rho\approx 10^{5-6}\
{\rm g\ cm^{-3}}$. The ashes from the burning form a degenerate liquid
ocean of unburned hydrogen and heavy nuclei. Hydrogen electron
captures occur once $E_F=1293\ {\rm keV}$, or $y\approx 10^{10}\ {\rm
g\ cm^{-2}}$, $\rho\approx 10^{7}\ {\rm g\ cm^{-3}}$. The thickness of
the electron capture transition layer is roughly a scale height. The
ocean crystallizes and forms the crust at $y\approx 10^{13}\ {\rm g\
cm^{-2}}$, $\rho\approx 10^{9-10}\ {\rm g\ cm^{-3}}$.  The ocean is
nearly isothermal beneath the electron capture transition layer with a
temperature of $5$-$7\times 10^8\ {\rm K}$.  For purposes of comparison,
the ocean thickness in relation to the radius is about the same as
that on the Earth.
\label{fig:scheme}}
\end{figure}

Figure \ref{fig:scheme} is a schematic of the NS atmosphere and
ocean. For accretion rates $\dot M < 10^{-8} M_\odot \ {\rm yr^{-1}}$,
the accreted hydrogen (H) and helium (He) rich material burns unstably
at a density of $\rho \lesssim 10^6 \ {\rm g \ cm^{-3}}$ and within a
few hours to days upon arrival on the star. The very high temperatures
reached ($T> 10^9 {\rm \ K}$) during the thermal instability produce
elements at and beyond the iron group. The isotopic mixture from this
burning is not well known, though it seems to always be the case that
a substantial amount of H (the residual mass fraction is typically
$X_r\sim 0.1$) remains unburned. The ashes from the burning accumulate
on the NS, undergo further compression and form a relativistically
degenerate ocean.  The hydrogen is eventually depleted by electron
captures at $\rho \approx 2\times 10^7\ {\rm g \ cm^{-3}}$, leading to
an abrupt rise in density at that location. At still higher densities,
$\rho \sim 10^9$--$10^{10} \ {\rm g \ cm^{-3}}$, the material crystallizes
and forms the crust. The thickness of the ocean is $\approx 10^4 \
{\rm cm} \ll R$.
There is no overt evidence for magnetic fields on these
NSs (i.e. none of them are persistent X-ray pulsars) and most
arguments about the nature of Type I X-ray bursts limit $B < 10^9 \
{\rm G}$ (Bildsten 1998a), weak enough not to affect the ocean g-modes
(BC95).

There has been prior work on g-modes in the upper atmosphere of the
NS, above and around the H/He burning region. McDermott \& Taam (1987)
calculated g-modes of a bursting atmosphere. SL96 calculated the
non-adiabatic mode structure for atmospheres accreting and burning in
steady-state and found that g-modes may be excited by the
$\epsilon$-mechanism (i.e. by the nuclear burning) when
$\dot{M}<10^{-10}M_\odot \ {\rm yr^{-1}}$. However, steady-state
burning does not occur at these low $\dot M$'s, and realistic
calculations for the time dependent NS atmosphere have yet to be
carried out.

Our focus in this paper is on the adiabatic mode structure in the deep
ocean underneath the H/He burning where the thermal time (hours to
days) is much longer than the mode period. The g-modes reside in the
ocean as they are effectively excluded from the crust by the restoring
force from the finite shear modulus there (McDermott et al. 1988;
BC95). If the NS is slowly rotating, the dispersion relation for these
modes is that for shallow water waves, $\omega^2\propto k^2$, where
$k=(l(l+1))^{1/2}/R$ is the transverse wavenumber and $l$ is the
angular eigenvalue. The transverse wavenumber changes if the star is
rapidly rotating, but this is easily calculable (BUC96) and it is
straightforward to find the g-mode spectrum from the non-rotating
calculations. 

The different sources of buoyancy yield a rich spectrum
of g-modes. The abrupt rise in density associated with the hydrogen
electron capture boundary layer supports a density discontinuity mode
of frequency $f_d\approx 35\ {\rm Hz} (X_r/0.1)^{1/2}$ for $l=1$ and
$R=10\ {\rm km}$ (equation [\ref{eq:fd}]). The internal buoyancy due
to the composition gradient within the electron capture boundary layer
creates a new spectrum of modes which are ``trapped''. Much of the
mode energy is confined to the boundary layer and our WKB estimate of
the mode frequency gives $f_{tr}\approx 8.5\ {\rm Hz}/n_{\rm tr}
(X_r/0.1)^{1/2}$ for $l=1$ and $R=10\ {\rm km}$ (equation
[\ref{eq:ftr}]).
 where $n_{\rm tr}$ is the number of nodes in the boundary layer.  
There is also a set of thermal g-modes (BC95) in the same frequency range as
the trapped modes, which are separated by the density
discontinuity into two distinct sets (see \S \ref{sec:thmodes}).

\subsection{Outline of This Paper} 

 We begin in \S 2 by explaining the physics of the deep ocean, the
electron captures that occur there and the subsequent neutron captures
by the heavy nuclei. We introduce a convenient fitting formula for the
electron capture rate and use it to generate a series of stellar
models for the subsequent seismic investigation. Section 3 summarizes
the equations for non-radial oscillations, discusses the various
sources of buoyancy in the deep ocean and our method for solving the
eigenvalue problem.  We present the full spectrum of g-modes for
isothermal oceans in \S 4.  There, we show both the modes due to the
electron capture discontinuity and the ``trapped'' modes, and find
that the thermal modes are split into two distinct sets by the density
discontinuity. We use the WKB approximation to obtain reliable
frequency estimates for these modes. We construct a few realistic
non-isothermal ocean models in \S 5 and show their g-mode spectra. We
summarize our results in \S 6 and discuss the effects of rapid
rotation and the important issues of mode excitation, damping and
transmittal of the signal to the observer.

\section{Electron Capture on Hydrogen in Accreting Neutron Stars}
\label{sec:layer} 

Rosenbluth et al. (1973) first noted that accreted matter would
undergo electron capture on hydrogen ($e^-+p\rightarrow n+\nu_e$) once
high enough densities were reached ($\rho>10^7\gcm$) for the electron
Fermi energy to exceed the 1293 keV threshold. Hydrogen electron
captures have since been studied in two very different contexts in
accreting neutron stars. The first is at very low accretion rates,
where the thermonuclear consumption of hydrogen (say via the CNO or pp
cycle) is so slow that the matter reaches the capture density prior to
having burned any hydrogen.  In this hydrogen-rich environment, the
neutrons created by this process are immediately captured by protons
(${\rm p} +{\rm n} \rightarrow {\rm D} + \gamma$).  Further reactions
burn this matter to at least helium, releasing 7 MeV per accreted
proton. Van Horn \& Hansen (1974) modeled transient X-ray sources as
the result of electron capture triggered thermonuclear flashes on low
mass ($M<0.15M_\odot$) neutron stars. A variant of this model was
later invoked as a possible source of local gamma-ray bursts by
Hameury et al. (1982), who found that the accretion rates where the
electron capture ignites a thermonuclear instability is comparable to
that expected from a lone neutron star steadily accreting from the
interstellar medium.

The second context is at the high accretion rates typical for the
brightest X-ray sources. In these stars, the matter undergoes
unstable H/He burning within a few hours to days of arrival on the NS,
long before reaching densities high enough for electron
capture. Provided $\dot M>\dot M_{c2} \approx 10^{-10} M_\odot \ {\rm
yr^{-1}}(Z_{\rm CNO}/0.01)^{1/2}$ (Bildsten 1998a), the H burning (via the
$\beta$-limited hot CNO cycle) is thermally stable while the H
accumulates.  For accretion rates in excess of
\begin{equation}\label{eq:mc1dot} 
\dot M_{c1}\approx 7\times 10^{-10}{M_\odot 
{\rm yr^{-1}}}\left(1.4 M_\odot\over M\right)^{2/9}
\left(R\over 10 \ {\rm km}\right)^{22/9}\left(Z_{\rm CNO}\over
0.01\right)^{13/18},
\end{equation} 
(Bildsten 1998a) the CNO cycle cannot consume all of the H before the
He unstably ignites (Lamb \& Lamb 1978; Taam \& Picklum 1979;
Fujimoto, Hanawa \& Miyaji 1981) in a local thermonuclear flash.  For
$\dot M_{c2} < \dot M < \dot M_{c1}$, the H is completely burned in a
stable manner and the accumulated pile of pure He unstably
ignites. There are many neutron stars accreting at rates comparable to
$\dot M_{c1}$.  However, the strong dependence of $\dot M_{c1}$ on
$Z_{CNO}$ and $R$ makes it difficult to know in which regime a
particular star is burning, especially if a large amount of spallation
occurs when the infalling matter decelerates at the surface (Bildsten,
Salpeter \& Wasserman 1992).\footnote{Many of the recent
interpretations of the kHz QPO's from these same objects imply that
the NS radius is inside the last stable orbit (Kaaret et al. 1997;
Zhang et al. 1997), resurrecting some of the original suggestions of
Kluzniak \& Wagoner (1985) and Kluzniak, Michelson \& Wagoner (1990).
If true, then spallation might reach the levels found by Bildsten et
al. (1992), as the flow can have a large radial component when it
finally reaches the surface.} The ashes from these two types of
unstable burning are quite different. In the pure He case, the ashes
are mostly $^{56}$Ni (Joss 1978; Ayasli \& Joss 1982) and there is
little to no hydrogen left at densities in excess of $10^6 \ {\rm g \
cm^{-3}}$. Things are different when the He ignites and burns in a
hydrogen-rich environment ($\dot M>\dot M_{c1}$), as we now discuss.

During the thermonuclear flash in the mixed H/He regime, hydrogen
burning is accelerated by freshly minted seed nuclei produced by He
burning, enhancing the local energy release and substantially
complicating the nuclear reaction chains. The very high temperatures
reached ($T\gtrsim 10^9 {\rm \ K}$) during the thermal instability
produce elements beyond the iron group (Hanawa, Sugimoto \& Hashimoto
1983; Wallace \& Woosley 1984; Schatz et al. 1997). The isotopic
mixture from this burning is still not well known, though everyone
agrees that a substantial amount of hydrogen (the residual mass
fraction is typically $X_r\sim 0.1$--$0.2$) remains unburned (Ayasli
\& Joss 1982; Woosley \& Weaver 1984; Taam et al. 1993; Schatz et
al. 1997). As the mix of H and extremely heavy elements accumulates
beneath the H/He burning layers, at temperatures in the range
$T\approx (2$--$5) \times 10^8 \ {\rm K}$, some proton captures by the
heavy nuclei occur (Schatz 1997, private communication). However, the
main proton consumption occurs when the matter has been compressed by
accretion to densities high enough for hydrogen electron capture
(Woosley \& Weaver 1984; Fushiki et al. 1992; Taam et al. 1996).
Electron capture onto the nuclei themselves occurs at much greater
densities and has been discussed previously by Sato (1979), Haensel \&
Zdunik (1990a, 1990b) and Blaes et al. (1990).

  In this section, we discuss the nuclear physics of the electron
capture transition layer in the ocean underneath the H/He burning
layer for those objects with $\dot M>\dot M_{c1}$.  We find that the
neutrons produced by the electron captures recombine with the heavy
nuclei, substantially changing the mix of heavy elements in the layer
and below.  Our present motivation is to calculate the internal buoyancy
associated with the composition gradient in the layer where electron
captures are occurring, and the resulting spectrum of g-modes.  We
find trapped g-modes and a density discontinuity g-mode that directly
depend on the internal buoyancy of the electron capture transition
layer, and thus only arise when hydrogen is present at these large
depths. The presence of such  g-modes would identify the star as being
in the $\dot M> \dot M_{c1}$ regime. 

\subsection{The Electron Capture Rate for Hydrogen} 

The electron capture rate for hydrogen (i.e. $e^-+p\rightarrow n
+\nu_e$) residing in an electron gas at temperature $T$ with Fermi
energy $E_F$ (including electron rest mass) is
\begin{equation}
R_{ec}(E_F,T)=\left({\ln2 \over ft}\right) I(E_F,T), 
\end{equation}
(Lang 1980) where $ft=1065\ {\rm s}$ as inferred from the measured
neutron half-life (Barnett et al. 1996) and the known $f$ value.  The
function $I$ is a dimensionless phase space integral over the electron
energy $E$,
\begin{equation}\label{eq:truerat}
I(E_F,T)={1\over (m_ec^2)^5}\int_Q^\infty{{E(E^2-m_e^2 c^4)^{1/2}(E-Q)^2
dE}\over {1+\exp[(E-E_F)/k_BT]}},
\end{equation}
where $Q=(m_n-m_p)c^2=1293.318 \ {\rm keV}$ is the energy threshold
for the reaction.  We have evaluated the integral in equation
(\ref{eq:truerat}) and agree with the tabulated rates of Oda et
al. (1994) to within 2\%. For the purposes of both analytical insight
and numerical speed, we now discuss a few limiting forms for the
capture rate and present a convenient fitting formula.

Much insight is gained if we assume relativistic electrons and neglect
the $m_e^2 c^4$ term in the integral $I$. There are then two limits to
discuss.  For $T=0$, captures only occur once the threshold is
exceeded. For values of the Fermi energy slightly above threshold
($E_F-Q\ll Q$), the integral becomes
\begin{equation}\label{eq:caprat}
I(E_F,T) \rightarrow {1\over 3}\left(Q\over m_ec^2\right)^5\left({{E_F-Q}\over
Q}\right)^3, \hskip 15 pt
{\rm for \ T=0 \ near \ threshold}.
\end{equation}
When $E_F\lesssim Q$, captures only occur if the electrons have a
finite temperature. The integral can be approximately found when
$Q-E_F\gg k_BT$, giving
\begin{equation}\label{eq:captemp}
I(E_F,T)\rightarrow {{2 Q^2 (k_BT)^3}\over{ (m_ec^2)^5}}\exp\left({E_F-Q}\over
k_B T\right), \hskip 15 pt
{\rm for } \ Q-E_F \gg k_BT, 
\end{equation}
so that most captured electrons are far out on the thermal tail of the
distribution. 

\begin{figure}[hbp]
\centering{\epsfig{file=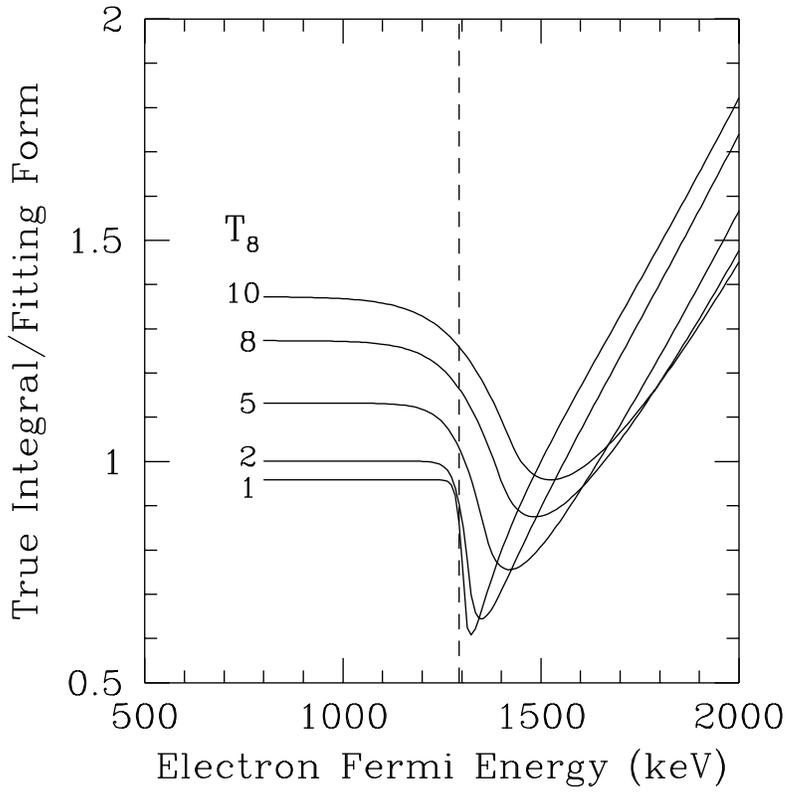,width=6 in}}
\caption{
The comparison of the exact integral to the analytic fit for the
hydrogen electron capture rates. The solid lines show the ratio of the
exact integral over the electron energy (eq. [\ref{eq:truerat}]) to
that determined by our simple analytic fit (eq. [\ref{eq:ratfit}]) for
temperatures of $T_8=1,2,5,8 $ and 10. These curves are tabulated and
later used to correct the simple formula.  The vertical dashed line
denotes where $E_F=Q$.
\label{fig:fits}}
\end{figure}

These limiting forms motivate a fitting formulae for the rates that we
now introduce. Since this might prove a convenient form for others to
use, we motivate our choice and discuss how to use it. It represents a
modified (and smooth) combination of equations (\ref{eq:caprat}) and
(\ref{eq:captemp}). The electron capture rate is temperature sensitive
even above threshold when $E_F-Q\sim k_BT$. We thus add a term
$\propto k_BT$ to $E_F$ in equation (\ref{eq:caprat}) which provides
some thermal ``fuzziness'' to the Fermi energy.  The proportionality
constant and Fermi energy to connect the modified equation
(\ref{eq:caprat}) with the unmodified equation (\ref{eq:captemp}) are
found by demanding that the rate and its first derivative with respect
to $E_F$ (at constant $T$) are continuous there. This results in the
form
\begin{equation}\label{eq:ratfit}
I\approx {Q^2\over (m_e c^2)^5 } \times \cases {
2(k_BT)^3 \exp[(E_F-Q)/k_BT] &  $ E_F<Q+k_BT \ln (9/2)$\cr 
(1/3)(E_F+(3-\ln 9/2)k_BT-Q)^3 & $E_F>Q+k_BT
\ln (9/2)$, \cr}
\end{equation}
which reproduces the electron capture rate within a factor of two over
a large range of temperatures and densities. Figure \ref{fig:fits}
shows the ratio of the true integral (eqn. [\ref{eq:truerat}]) to that
given by equation (\ref{eq:ratfit}) for a range of $E_F$ and
temperatures of $T_8=1,2,5,8$ and 10. Clearly the fit does rather well
at representing a function that is changing by orders of magnitude in
this range. The growing deviation at $E_F\gg Q$ arises from our
neglect of the next order term in equation (\ref{eq:caprat}), which is
$\propto (E_F-Q)^4$.

We implement this fit by first tabulating (as a function of $E_F$ and
$T$) the ratio of the real answer to the fit.  Then, for any $E_F$ and
$T$, we use equation (\ref{eq:ratfit}) to evaluate $I$ and then find
the correction via interpolation within this pre-made table. This
proves to be quite accurate with a relatively small table, as the
fitting formula takes care of most of the strong variations in $I$.

\begin{figure}[hbp]
\centering{\epsfig{file=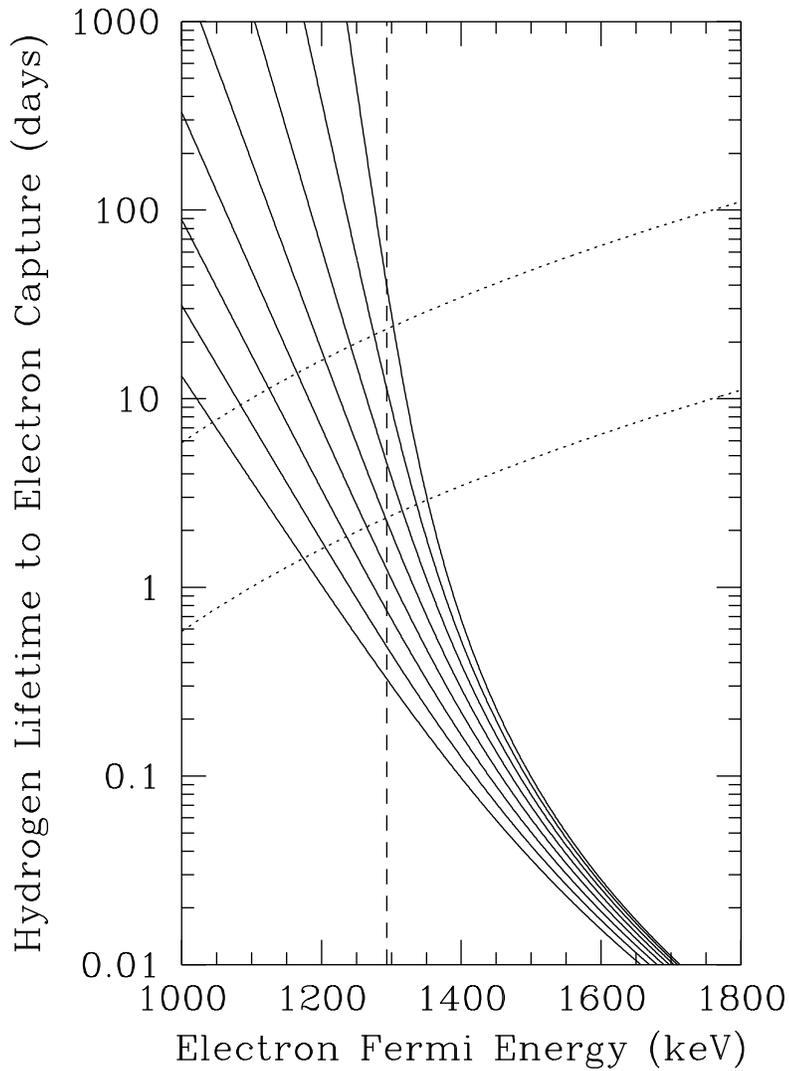,width=6 in}}
\caption{
The proton lifetime to electron capture as a function of the
electron Fermi energy.  The solid lines show the lifetime of a proton
($1/R_{ec}$) as a function of the electron Fermi energy
for a range of temperatures. Starting from the uppermost solid curve, we
display the lifetime for $T_8=2,3,4,5,6,7,8 $ and 9.  The vertical
dashed line denotes where $E_F=Q$. The dotted curves are the rough
measure of the time it takes matter to reach that depth for 
$\dot m=\dot m_{\rm Edd}$ (lower dotted) and 
$\dot m=0.1\dot m_{\rm Edd}$ (upper dotted). 
\label{fig:timecomp}}
\end{figure}

We describe the structure of the hydrogen electron capture transition
layer in \S \ref{sec:laystr}, but we can obtain a rough picture of
where the electron captures occur by noting that in steady state, the
hydrogen depletes when the lifetime to electron capture, $t_{ec}\equiv
1/R_{ec}$, becomes comparable to the time to reach that depth,
$t_{accr}\equiv y/\dot m$, where $\dot m$ is the local accretion rate
(mass accreted per unit time per unit area) and $y=p/g$ is the column
depth.  The local Eddington accretion rate is
\begin{equation}
\dot m_{Edd}={\mu_em_pc\over\sigma_TR}=7.5\times 10^4\ 
\mu_e\left({R\over 10\ {\rm km}}\right)
{\rm\ g\ cm^{-2}\ s^{-1}},
\end{equation}
where $\sigma_T$ is the Thomson cross-section and $\mu_e$ is the mean
molecular weight per electron. For simplicity, we take $\dot m_{Edd}$
to be the value for $\mu_e=1$ and $R=10\ {\rm km}$.  Figure
\ref{fig:timecomp} shows $t_{ec}$ as a function of $E_F$ for
many different temperatures. From top to bottom, the solid lines are
for $T_8=2,3,4,5,6,7,8,$ and $9$. The vertical dashed line displays
$E_F=Q$. The lower (upper) dotted line shows $t_{\rm accr}$ for $\dot
m=\dot m_{\rm Edd}$ ($\dot m=0.1\dot m_{\rm Edd}$), where, for
simplicity, we have assumed the degenerate electrons provide all
the pressure.  It is clear that for $T_8\gtrsim 6 $ at $\dot m=\dot
m_{\rm Edd}$ and $T_8\gtrsim 3$ at $\dot m=0.1
\dot m_{\rm Edd}$, substantial pre-threshold capture occurs, i.e. many
protons are consumed while $E_F<Q$. We discuss this further in \S 
\ref{sec:laystr}. 

\subsection {The Fate of the Free Neutrons}

In the hydrogen-rich environment, the neutrons produced by the
electron captures rapidly thermalize via elastic scattering with
protons and undergo a radiative capture (free decay is irrelevant for
these high densities, as the radiative capture times are $\ll
10^{-10}$ seconds).  Woosley \& Weaver (1984) and Taam et al. (1996)
assumed that the neutrons would capture onto protons, starting a
chain of rapid nuclear reactions up to the Fe group. We find that this
is not the case.  The reaction rate for capture onto protons (${\rm n
\ +\ p} \rightarrow {\rm D}\ +\ \gamma$) is $\langle \sigma
v\rangle=7.3\times 10^{-20} \ {\rm cm^3 \ s^{-1}}$ (Caughlan \& Fowler
1988), whereas typical rates for neutron captures onto heavy nuclei
are $10^{-17}$--$10^{-19}\ {\rm cm^3 \ s^{-1}}$, nearly two orders of
magnitude larger. We thus find that, for the hydrogen mass fractions
we consider, most neutrons capture onto the heavy nuclei, releasing
the typical 7--8 MeV binding energy of a nucleon in a heavy nucleus.
For example, from Cowan, Thielemann \& Truran (1991), the neutron
capture rate for $^{56}$Fe is $5.8\times 10^{-18}{\rm cm^3 \ s^{-1}}$
(at 30 keV), so that neutrons will capture preferentially onto iron
rather than protons when their ratio by number is $\lesssim 80$,
requiring $X\lesssim 0.58$. Similar conclusions apply to most nuclei
we could consider.

The neutron captures substantially modify the isotopic mix of heavy
elements. At a particular depth in the transition layer, those
elements with larger neutron capture cross sections grow more and more
neutron rich until they become $\beta$-unstable on a timescale
comparable to the flow time across the region (i.e. days, see Figure
\ref{fig:timecomp}). Since the composition of the ashes from
the H/He burning is still unknown, we adopt a simple model in which we
assume that prior to electron capture the gas consists of protons,
electrons and a {\it single} species of nucleus of mass $Am_p$ and charge
$Ze$.  The mass fraction of protons $X$ is given by $\rho_p=X\rho$, so
that the density of nuclei is $\rho_N=(1-X)\rho$ and the local number
ratio of protons to nuclei is $n_p/n_N=XA/(1-X)$. The mean molecular
weight per electron for this mixture is 
\begin{equation} 
{n_em_p\over \rho}\equiv {1\over \mu_e}=X+{Z\over A}\left(1-X\right),
\end{equation}
where $n_e$ is the electron number density. We assume that only one
nuclear species is present at a particular depth. Given the starting
values $X_r$ and $A_i$, the value of $X$ at a particular depth fixes
the mass of the nucleus $A$ due to baryon conservation, $A/(1-X)={\rm
const}=A_i/(1-X_r)$.  Thus $A$, in some sense, represents the average
weight of the nuclei present, and is a continuous variable. The models
we discuss throughout this paper are for the ``neutron-rich'' case in
which we neglect $\beta$ decays and keep $Z$ fixed and let $A$ grow
without bound. We have also constructed models in which the nuclei
follow the valley of stability, i.e. at a particular depth we choose
the most stable $Z$ corresponding to the local value of $A$. Clearly,
the real composition of the layer lies somewhere between these two
cases. However, we find that (except for the density discontinuity
mode, see \S \ref{sec:dismodes}) the adiabatic seismology depends very
weakly on the exact nuclear physics in the layer, and so we adopt the
``neutron rich'' models for simplicity. We have yet to fully explore
the rich nuclear physics of this transition layer.

\subsection{The Structure of the Hydrogen Electron Capture Boundary Layer} 
\label{sec:laystr} 

 At the accretion rates of interest for the bright X-ray sources, the
downward diffusive drift speed of a nucleus in the hydrogen-rich
region is much less than the downward flow speed from accretion, $\vec
v=-v \hat z=-\dot m \hat z/\rho$. The nuclei thus do not have time to
settle out (gravitationally separate) from the hydrogen before the
electron captures occur. For example (using equation 4.3 from
Bildsten, Salpeter \& Wasserman 1993), for temperatures in excess of
$5\times 10^8 \ {\rm K}$, there is no relative diffusion of protons
and heavy ions provided $\dot M > 10^{-10} M_\odot {\rm
yr^{-1}}$. Thus there are no relative separations in the continuity
equations (Brown \& Bildsten 1998), and we construct models of the
hydrogen electron capture boundary layer by integrating the continuity
equation for protons,
\begin{equation}
{\partial n_p \over
\partial t}+\vec \nabla \cdot (n_p \vec v) = -n_p R_{ec}. 
\end{equation} 
We rewrite this equation in terms of the column density, $y=\int
\rho(z) dz=p/g$, and the hydrogen mass fraction $X$, giving
\begin{equation}
{\partial X \over \partial t }+ \dot m {\partial X\over
\partial y}= - X R_{ec},
\end{equation} 
where we have used the mass continuity equation.  In a  steady-state, the
electron captures are balanced by the accretion flow, so that
\begin{equation}\label{eq:simpcont}
\dot m {dX\over dy}=-X R_{ec}.
\end{equation}
Since the temperature gradient across the transition layer is small
(see \S \ref{sec:nonthmod}), we first presume the layer is isothermal
in our calculations. In this case, equation (\ref{eq:simpcont}) gives
a complete description of the transition layer. We use these isothermal
models in \S \ref{sec:modes} to illustrate the spectrum of ocean
g-modes. We discuss realistic non-isothermal models of the whole ocean
in \S \ref{sec:models}, in which the heat equation must also be
integrated.  A discussion of the equation of state and the
microphysics of the ocean can be found in BC95. The temperature of the
ocean is $T_8\equiv (T/10^8\ {\rm K})\approx 5$--$7$.

We now make a simple model of the hydrogen electron capture layer at
zero temperature. In this case no electron capture occurs until
$E_F>Q$. We neglect the electron rest mass, assume the boundary layer
is thin, write the pressure as $p=gy=n_eE_F/4$, and use the capture
rate of equation (\ref{eq:caprat}). We then integrate equation
(\ref{eq:simpcont}) analytically, obtaining $X(E_F)=X(Q)\exp(-B)$,
where
\begin{equation}\label{eq:B}
B\equiv \int_Q^{E_F} {R_{ec}\over\dot m}dy 
\approx
{2\pi\over 9(hc)^3}{1\over\dot{m}g}{\ln 2\over ft}
\left(\frac{Q}{m_ec^2}\right)^5(E_F-Q)^4.
\end{equation}
If we measure the characteristic ``thickness'', $\Delta E_F=E_F-Q$, of
the electron capture layer by the depth at which the hydrogen is 90\%
depleted, we find
\begin{equation}\label{eq:thickness}
\Delta E_F=172\keV\left(\frac{\dot{m}}{\dot{m}_{Edd}}\right)^{1/4}
\left(\frac{g}{2\times 10^{14}\ {\rm cm\ s^{-2}}}\right)^{1/4}
\left({Q\over 1293\ {\rm keV}}\right)^{-5/4}
\left({ft\over 1065\ {\rm s}}\right)^{1/4}.
\end{equation}
Integrating the equation of hydrostatic equilibrium, $dz=-dy/\rho$
gives the corresponding physical thickness of the layer,
\begin{equation}\label{eq:physthick}
\Delta z={826\cm\over\Sigma}
\left(\frac{\dot{m}}{\dot{m}_{Edd}}\right)^{1/4}
\left(\frac{g}{2\times 10^{14}\ {\rm cm\ s^{-2}}}\right)^{1/4}
\left({Q\over 1293\ {\rm keV}}\right)^{-5/4}
\left({ft\over 1065\ {\rm s}}\right)^{1/4}
, 
\end{equation}
where
\begin{equation}
{1\over\Sigma}\approx\frac{Z}{A_i}(1-X_r)+{(-3/4)!\over 5}X_r
\approx\left({1\over\mu_e}\right)_i,
\end{equation}
and the subscript $i$ denotes initial values, $X_r$ is the initial
hydrogen mass fraction, $(-3/4)!=3.63$, and we have assumed that
$Z$ is constant through the layer. We include the scalings with $ft$
and $Q$ in these formulae as they are equally applicable to electron
capture layers associated with nuclei at much larger depths. Numerical
integrations of equation (\ref{eq:simpcont}) with the full $T=0$
equation of state for the electrons and the exact expression for the
electron capture rate (equation [\ref{eq:truerat}]) find good
agreement with both the scalings and prefactors given in equations
(\ref{eq:thickness}) and (\ref{eq:physthick}) ($\lesssim 5\%$ for
$\dot{m}\lesssim 10\dot m_{\rm Edd}$ after which the term $\propto
(E_F-Q)^4$ neglected in equation (\ref{eq:caprat}) becomes important).

\begin{figure}[hbp]
\centering{\epsfig{file=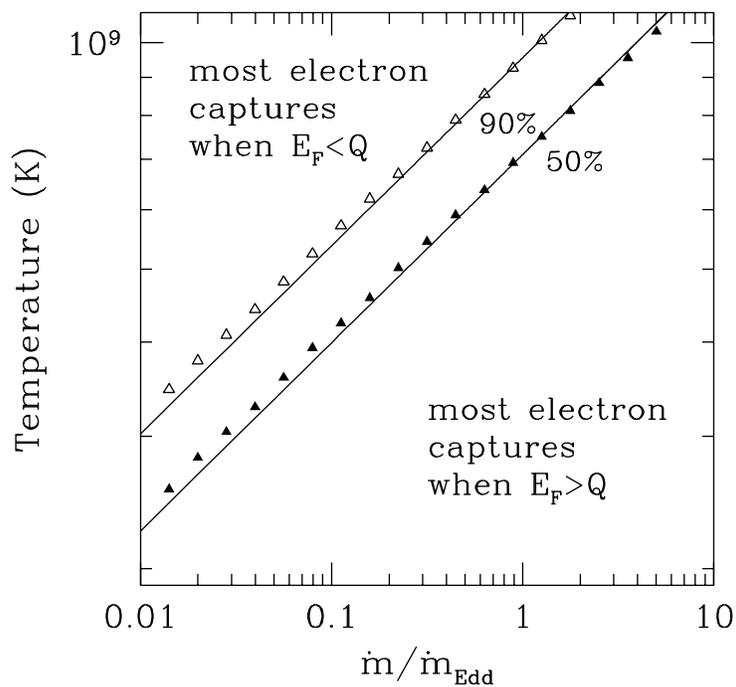,width=4 in}}
\caption{
Pre-threshold vs. post-threshold electron capture on hydrogen.  The
$\dot{m}$--$T$ plane is divided into two regions: at high $T$ or low
$\dot m$, most captures happen pre-threshold ($E_F<Q$); at low $T$ or
high $\dot m$, most captures happen post-threshold ($E_F>Q$).
The filled (open) triangles show points at which 50\% (90\%) of
protons have captured pre-threshold. The solid lines show the
$T\propto\dot m^{1/4}$ scaling (equation [\ref{eq:Tmdot}]).
\label{fig:Tmdot}}
\end{figure}

As shown in Figure \ref{fig:timecomp}, at high enough temperatures
significant electron capture occurs pre-threshold. For a particular
$\dot m$, there is a critical temperature $T_c$ above which this
occurs, roughly given by $k_BT_c\sim \Delta E_F$ from equation
(\ref{eq:thickness}). We accurately estimate $T_c$ by integrating the
continuity equation (\ref{eq:simpcont}) with the {\it pre-threshold}
expression for the electron capture rate (equation
[\ref{eq:captemp}]). Again neglecting the electron mass, we find that
when
\begin{equation}\label{eq:Tmdot}
T_8> T_c=7.1
\left(\frac{\dot{m}}{\dot{m}_{Edd}}\right)^{1/4}
\left(\frac{g}{2\times 10^{14}\cm\secd^{-2}}\right)^{1/4}
\left({Q\over 1293\ {\rm keV}}\right)^{-5/4}
\left({ft\over 1065\ {\rm s}}\right)^{1/4}
, 
\end{equation}
more than $50\%$ of the protons have captured electrons pre-threshold. The
prefactor in this expression is taken from numerical integration of
equation (\ref{eq:simpcont}) for an isothermal layer with the full
electron capture rate (eq. [\ref{eq:truerat}]) and equation
of state.  Figure \ref{fig:Tmdot} shows the critical lines in the
$\dot{m}$--$T$ plane for which 50\% and 90\% of the protons capture
pre-threshold. There is excellent agreement with the analytic
estimate.

\section{Adiabatic Oscillations in the Ocean}
\label{sec:buoyancy}

 Our present focus is on those g-modes that reside predominantly in the
liquid ocean that lies beneath the H/He burning layer. The
oscillations are adiabatic there since the local thermal time (hours
to days) is much longer than the g-mode period.  We first discuss the
different sources of buoyancy in the ocean and then describe the
adiabatic perturbation equations and our method for solving them. 

\subsection{Sources of Buoyancy}

The g-mode frequencies are set by the local buoyancy force, which
results from the density contrast between a displaced fluid element
and the background fluid.  There are two contributions to this density
contrast.  The first is thermal buoyancy, which arises because the
change in density of the fluid element as it is adiabatically
displaced differs from the change in the background density due to the
non-isentropic density gradient. The second contribution is from
composition gradients.  For example, in the hydrogen electron capture
layer the timescale for electron capture is much greater than the mode
period, so that the perturbed fluid element's composition is fixed
during the pulsation.  Because of the large composition gradient in
the layer, the perturbed element is heavier than the surroundings 
and is forced back. 

\begin{figure}[hbp]
\centering{\epsfig{file=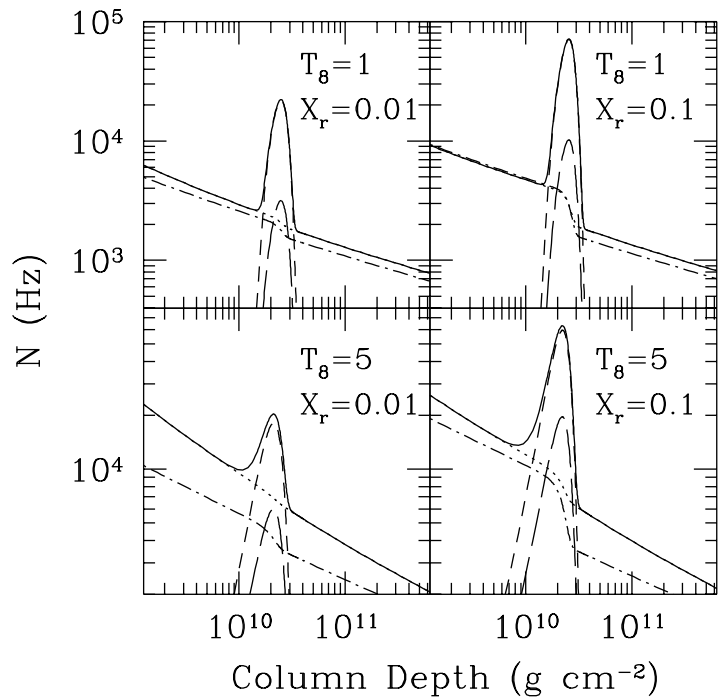,width=4 in}}
\caption{
The Brunt-V\"ais\"al\"a frequency for isothermal models accreting at
$\dot m=\dot m_{\rm Edd}$. The solid line shows $N$, and the different
lines show individual contributions. The thermal buoyancy ({\it dotted
line}) dominates above and below the electron capture boundary, while
the buoyancy due to the $\mu_e$ gradient ({\it short dashed line})
dominates in the boundary layer. The $\mu_i$ buoyancy ({\it long
dashed line}) is about the same as the thermal buoyancy in the layer.
The dot-dashed line shows the analytic estimate of the thermal
buoyancy (equation (\ref{eq:N2therm})), which performs best at low
temperatures.
\label{fig:N2}}
\end{figure}

The local buoyancy is measured by the Brunt-V\"ais\"al\"a frequency
$N$, given by
\begin{equation}\label{eq:N2A}
N^2=-g{\cal A}=-g\left(\dlnz{\rho}-\gamu\dlnz{p}\right)
=-g\left(\dlnz{\rho}+{1\over\Gamma_1h}\right)
\end{equation}
where $\Gamma_1\equiv (\partial\ln p/\partial\ln\rho )_s$ is the
adiabatic index and $\cal A$ is the convective discriminant.  
For a mixture of electrons and ions, in
which the pressure is a function of density $\rho$, mean molecular
weight per electron $\mu_e$ and mean molecular weight per ion $\mu_i$,
we can rewrite $\cal A$ as
\begin{equation}
{\cal A}={d\ln p\over
dz}\left({1\over\chi_\rho}-{1\over\Gamma_1}\right)
-{\chi_T\over\chi_\rho}{d\ln T\over dz}
-{\chi_{\mu_e}\over\chi_\rho}{d\ln\mu_e\over dz}
-{\chi_{\mu_i}\over\chi_\rho}{d\ln\mu_i\over dz},
\end{equation}
where $\chi_{\rm Q}=(\partial\ln p/\partial\ln {\rm Q})$ with the
other independent variables held constant.  Using the thermodynamic
identities $\Gamma_1=(\Gamma_3-1)\chi_T+\chi_\rho$, and
$\nabla_{ad}\equiv (\partial\ln T/\partial\ln
p)_s=\Gamma_3-1/\Gamma_1$ (Cox \& Giuli 1968), we obtain
\begin{equation}\label{eq:N2}
{N^2h\over g}={\chi_T\over\chi_\rho}\left[\nabla_{ad}-\left({d\ln T \over
d\ln p}\right)_\ast\right]
-{\chi_{\mu_e}\over\chi_\rho}\left({d\ln\mu_e\over d\ln p}
\right)_\ast
-{\chi_{\mu_i}\over\chi_\rho}\left({d\ln\mu_i\over d\ln p}
\right)_\ast,
\end{equation}
where the subscript $\ast$ refers to the stellar model and $h=p/\rho
g$ is the local pressure scale height. The first term in this equation
is the thermal buoyancy and the remaining terms are contributions from
composition gradients.  Figure \ref{fig:N2} shows the different
contributions to $N$ in isothermal models of the ocean. In the
electron capture layer, the buoyancy due to the $\mu_e$ gradient
dominates the other contributions, elsewhere the thermal buoyancy is
most important. The buoyancy due to the $\mu_i$ gradient is of order
the thermal buoyancy in the layer; despite $\chi_{\mu_i}$ being small
(the pressure is insensitive to $\mu_i$) the change in $\mu_i$ is
typically large across the layer, $\Delta\mu_i/\mu_i\approx 6$.  We
use isothermal models here for illustrative purposes. In reality, the
ocean has a slight temperature gradient (see \S
\ref{sec:nonthmod}), which changes the thermal buoyancy. However, 
the electron capture transition layer is typically thin enough so that
it can be described by a single temperature.

We now make an analytic estimate of the thermal buoyancy in the
isothermal case under some simplifying assumptions. First, we use the
relation $\nabla_{ad}=\chi_T p/\rho T c_V\Gamma_1$ to write the
thermal buoyancy as
\begin{equation} 
N^2_{\rm th}=
{\chi_T^2\over\Gamma_1\chi_\rho}{g^2\over c_V T},
\end{equation}
where $c_V$ is the heat capacity at constant volume.  There are two
contributions to $\chi_T$. At low temperatures, the electron
contribution is small, so that $\chi_T$ is dominated by the
contribution from the ions, $\chi_T\rightarrow \rho k_B T/\mu_i m_p p$
as $T\rightarrow 0$ (see Hansen \& Kawaler 1994 for a useful
discussion).  The heat capacity is
\begin{equation}\label{eq:cv}
c_V={k_B\over\mu_im_p}\left({3\over 2}+\pi^2{\mu_i\over\mu_e}{k_BT\over
E_F}\right),
\end{equation}
where the first (second) term is the contribution from the ions
(electrons). For this estimate, we ignore the electron
contribution and treat the ions as an ideal gas, neglecting Coulomb
corrections to the equation of state.  Our detailed numerical
calculations include all of these effects.  Taking
$\Gamma_1\approx\chi_\rho\approx 4/3$ for relativistic electrons, the
final result is (compare BC95 equation 3.9)
\begin{equation}\label{eq:N2therm}
N^2_{\rm th} h^2 \mu_i = {3k_BT\over 8m_p}. 
\end{equation}
This estimate is shown as the dot-dashed line in Figure \ref{fig:N2}.
It performs well at low temperatures, but underestimates the thermal
buoyancy at higher temperatures, where the contribution from the
electron entropy becomes non-negligible. The non-ideal gas corrections
for the ions (for which we use the parametrization of Farouki \&
Hamaguchi 1993) have little effect on $N$.

At zero temperature, the thermal buoyancy vanishes and all buoyancy
comes from the $\mu_e$ gradient. The pressure is entirely from the
degenerate electrons, giving $\Gamma_1=\chi_\rho=-\chi_{\mu_e}$ and
$\chi_{\mu_i}=0$, so that $N^2=-g\ d\ln\mu_e/dz.$ This form of
buoyancy has been studied before in the core of a neutron star where
the density dependent $\beta$-equilibrium means that $\mu_e$ changes
with position. Reisenegger \& Goldreich (1992) found a new class of
core g-modes in the $\lesssim 100 \Hz$ range from this effect. These
modes do not penetrate far into the crust and so most likely have
little impact in the neutron star ocean which we are studying. 

\subsection{Solving the Adiabatic Mode Equations}

The equations describing linear adiabatic perturbations of the thin ($h\ll
R$) ocean are 
\begin{equation}\label{eq:mode1}
\frac{d\xi_z}{dz}=\dpp\left(\frac{ghk^2}{\omega^2}-
\gamu\right)+\frac{\xi_z}{\Gamma_1h}
\end{equation}
\begin{equation}\label{eq:mode2}
\frac{d}{dz}\dpp=\dpp\frac{1}{h}\left(1-\gamu\right)+
\frac{\xi_z}{h}\left(\frac{\omega^2}{g}-\frac{N^2}{g}\right),
\end{equation}
where $\xi_z$ is the radial displacement of the perturbed fluid
element and $\delta p$ is the Eulerian pressure perturbation.  We
follow BC95 and use the notation $\delta$ to describe an Eulerian
perturbation and $\Delta$ to describe a Lagrangian perturbation,
$\Delta\equiv\delta+{\vec{\xi}\cdot \vec{\nabla}}$.  We adopt the
Cowling approximation, in which we neglect perturbations of the local
gravity. We use Newtonian gravity and take $g\equiv
GM/R^2$.
These
equations are in the plane-parallel limit since the depth of the ocean
is much less than the stellar radius. The horizontal wavenumber is
then $k=\sqrt{l(l+1)}/R$ for a slowly rotating star.  Bildsten et
al. (1996) showed that rapid rotation simply modifies this angular
eigenvalue, allowing our calculations to be easily extended to the
case of rapid rotation ($f_s\gg$ mode frequency).

Solving the adiabatic mode equations requires $h$, $\Gamma_1$ and $N$
throughout the ocean. BC95 showed that the shear modulus of the crust
effectively excludes g-modes because they have large transverse
shearing motions. We thus demand that the radial displacement
vanishes, $\xi_z=0$, at the ocean floor where the ions
crystallize. This is defined by $\Gamma =\Gamma_m\approx 173$ (Farouki
\& Hamaguchi 1993), where
\begin{equation}\label{eq:gamma}
\Gamma={Z^2e^2\over akT}=113\ {\rho_7^{1/3}\over T_8}
\left({Z\over 30}\right)^{5/3}
\left({2Z\over A}\right)^{1/3}
\end{equation}
measures the importance of Coulomb effects. Here $a$ is the average
ion spacing, $4\pi a^3n_i/3=1$, and we presume for simplicity that
only a single ion species is present.  Assuming relativistic electrons
and using the $T=0$ Fermi energy (very good approximations at the
crystallization depth), we find that the crust starts at a column
density
\begin{equation}\label{eq:crystal_y}
y_{cr}\approx 1.85\times 10^{13}\ {\rm g\ cm^{-2}} 
\left({T_8\over 5}\right)^{4}
\left({g\over 2\times 10^{14}\ {\rm cm\ s^{-2}}}\right)^{-1}
\left({Z\over 30}\right)^{-20/3}
\left({\Gamma_m\over 173}\right)^{4}.
\end{equation}
The ocean can crystallize before hydrogen electron captures occur
only if $T_8<1\ (Z/30)^{5/3}$, an unlikely circumstance at the accretion
rates we are considering.

We adopt a ``shooting'' method for solving the eigenvalue problem. For
a particular trial frequency, we integrate from the surface of the
ocean to the top of the crust. The boundary condition at the ocean
floor, $\xi_z=0$, is only satisfied when the trial frequency is an
eigenvalue of the equations. We start our integrations at an arbitrary
place (column depth $y_t=10^9\ {\rm g\ cm^{-2}}$) beneath the H/He
burning layer. We apply the boundary condition there that the
Lagrangian pressure perturbation vanishes ($\Delta p=0$). Although not
strictly correct, this boundary condition is reasonable for the low
frequency g-modes we discuss, and we find that the mode frequencies
are insensitive to changes in the upper boundary condition or starting
point for the integrations.\footnote{ The g-mode frequencies are
insensitive to details of the upper boundary, despite there being some
mode amplitude there, because they are mainly determined by buoyancy
deeper in the ocean.  The exceptions are the upper thermal g-modes,
and at high temperatures the trapped modes, whose frequencies depend
on the thermal buoyancy above the electron capture layer. For example,
changing $y_t$ from $10^9\ {\rm g\ cm^{-2}}$ to $5\times 10^8\ {\rm g\
cm^{-2}}$ changes the discontinuity and lower thermal mode frequencies
by $1\%$, the trapped mode frequencies by up to $10\%$ depending on
the temperature, and the upper thermal mode frequencies by $20\%$. We
discuss the effect on the upper thermal modes and trapped modes of
extending our calculations into the upper atmosphere in \S
\ref{sec:trapmodes} and \S
\ref{sec:thmodes}.}

\section{The Spectrum of Ocean g-Modes}
\label{sec:modes}

The distinct sources of buoyancy in the region beneath the H/He
burning yields an extremely rich spectrum of g-modes. The abrupt rise
in density from the hydrogen electron captures supports a single
density discontinuity mode, and the internal buoyancy due to the
composition gradient creates a new spectrum of modes which are
``trapped'' in the transition layer. There is also a set of thermal
modes, as originally described by BC95, which we now find are
separated by the density discontinuity and confined to either the
upper or lower parts of the ocean.

We use isothermal models with $A_i=60$ in this section in order to
illustrate and explain the physics of the different types of ocean
g-modes. The large amount of overlapping frequencies for these modes
means that we must first understand them separately before jumping in
to the realistic models that include temperature gradients. Our
description of the mode spectrum of non-isothermal models in \S
\ref{sec:models} is then much simpler.  These stars also have a
surface wave or f-mode and acoustic waves or p-modes. These modes,
which have frequencies $\gtrsim 10\ {\rm kHz}$, penetrate into the
crust because they have little transverse shear and are thus not well
represented by our thin ocean models and so we do not discuss them further.

\subsection{Discontinuity Modes}
\label{sec:dismodes} 

 As Finn (1987) noted in his original study of discontinuity modes
in isolated neutron stars, we gain some intuition by first considering
waves at the interface of two {\it constant density, incompressible}
fluids. This case can be solved analytically. Since the discontinuity
lies closer to the surface of the ocean than the floor, we take the
upper fluid (density $\rho_+$) to have a finite extent $H$ and the
lower fluid (density $\rho_-$) to extend to infinite depth.  In the
shallow water wave limit ($kH\ll 1$) the discontinuity mode frequency
is then 
\begin{equation}\label{eq:discincomp}
\omega^2_d=gk^2H\frac{\rho_--\rho_+}{\rho_-}.
\end{equation}
We apply this by ignoring the finite thickness of the hydrogen
electron capture layer for now (see \S \ref{sec:discfinite} for finite
thickness models), and considering a simple model, a $T=0$ ocean with
a discontinuity in $\rho$ and $\mu_e$ at the place where $E_F=Q$. The
buoyancy then vanishes everywhere, except at the discontinuity. To
solve this problem, we modified our shooting scheme and applied the
matching conditions $\xi_{z+}=\xi_{z-}$ and $\Delta p_+=\Delta p_-$
across the discontinuity (Finn 1987; McDermott 1990). In these
equations, a $+$ ($-$) refers to a quantity in the upper (lower) part
of the ocean.  These conditions are physically motivated; $\xi_z$ must
be continuous at the discontinuity because of the continuity equation,
while $\Delta p$ must be continuous to avoid infinite acceleration of
a fluid element. There is no restriction on the transverse
displacement $\xi_\perp$, and indeed we find a large jump in
$\xi_\perp$ across the discontinuity.

 Equation (\ref{eq:discincomp}) motivates a frequency estimate of the
density discontinuity mode associated with the electron capture layer.
The length scale governing the behavior of the eigenfunction in the
{\it compressible} case is the pressure scale height just above the
discontinuity
\begin{equation}\label{eq:scaleheight}
h_+=\frac{1560\cm}{\mu_{e+}}
\left({\EF\over 1293\keV}\right)
\left({g\over 2\times 10^{14}\ {\rm cm\ s^{-2}}}\right)^{-1},
\end{equation}
where we neglected the electron mass and assumed relativistic
degeneracy.  A natural guess for the discontinuity mode frequency is
then 
\begin{equation}\label{eq:fdguess}
\omega^2_d\approx gk^2h_+\frac{\rho_--\rho_+}{\rho_-}, 
\end{equation}
and at fixed pressure $(\rho/\mu_e)_+=(\rho/\mu_e)_-$, so that
(at T=0)
\begin{equation}\label{eq:fd2}
f_d=111\Hz\left[\left(\frac{1}{\mu_e}\right)_+-\left(
\frac{1}{\mu_e}\right)_-\right]^{1/2}
\left({10 \ {\rm km}\over R}\right)
\left({l(l+1)\over 2}\right)^{1/2},
\end{equation}
or for our case of a single ion species,
\begin{equation}\label{eq:fd}
f_d=35\Hz
\left({X_r\over 0.1}\right)^{1/2}
\left(1-\frac{\Delta Z}{\Delta
A}\right)^{1/2}
\left({10 \ {\rm km}\over R}\right)
\left({l(l+1)\over 2}\right)^{1/2},
\end{equation}
where $\Delta Z$ and $\Delta A$ are the changes in the charge and mass
of the nuclei from one side of the discontinuity to the other. The
prefactors in equations (\ref{eq:fd2}) and (\ref{eq:fd}) are from our
numerical calculations. The scalings found are as predicted by
equation (\ref{eq:fdguess}). The mode frequency is insensitive to the
position of the surface or floor of the ocean as long as they lie more
than a few scale heights away from the discontinuity. This is as we
expect in the spirit of our guess above; it is the shortest length
scale near the discontinuity which dominates the mode eigenfunctions
and sets the mode frequency in the limit of $kH\ll 1$.

\begin{figure}[hbp]
\centering{\epsfig{file=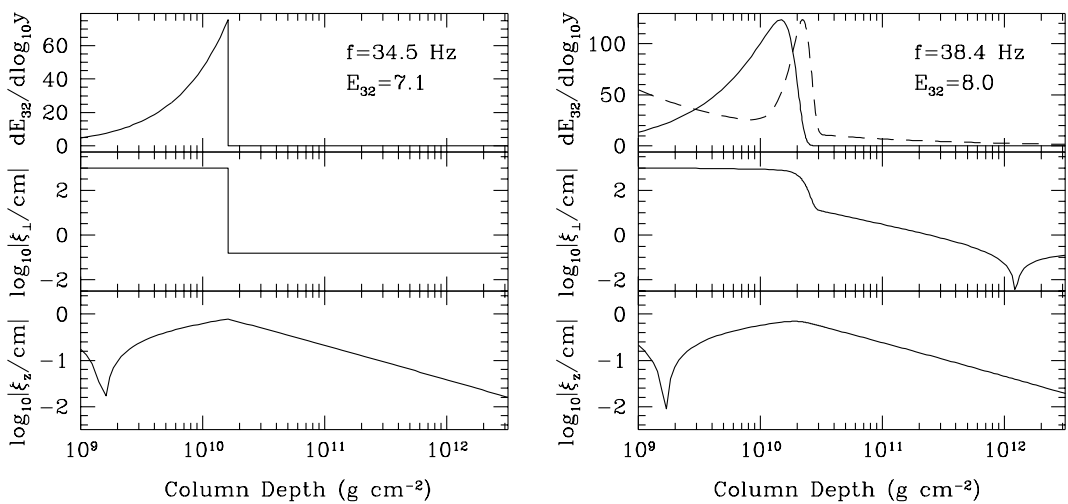,width=\hsize}}
\caption{
Eigenfunctions of the $l=1$ density discontinuity modes. The left
panel shows the density discontinuity mode for a $T=0$ ocean with
$X_r=0.1$. The right panel is for an isothermal ocean with $\dot
m=\dot m_{Edd}$, $T_8=5$ and $X_r=0.1$. The middle (bottom) panels
show the transverse (radial) displacement as a function of column
depth. The top panels show the mode energy density
$dE_{32}/d\log_{10}y$ ({\it solid line}) and Brunt V\"ais\"al\"a frequency
$N$ on a linear scale ({\it dashed line}). The displacements shown are
absolute values, so zero-crossings appear as cusps. We begin our
integrations at a column depth $y=10^9\ {\rm g\ cm^{-2}}$ and
integrate to the floor of the ocean.  In all our plots, we arbitrarily
normalize the transverse displacement to $\xi_\perp=10^3\cm$ at the
top of the ocean. The mode energy, $E_{mode}$, is measured in units of
$E_{32}\equiv E_{mode}/10^{32}\ {\rm erg}$.
\label{fig:disc}}
\end{figure}

\begin{figure}[hbp]
\centering{\epsfig{file=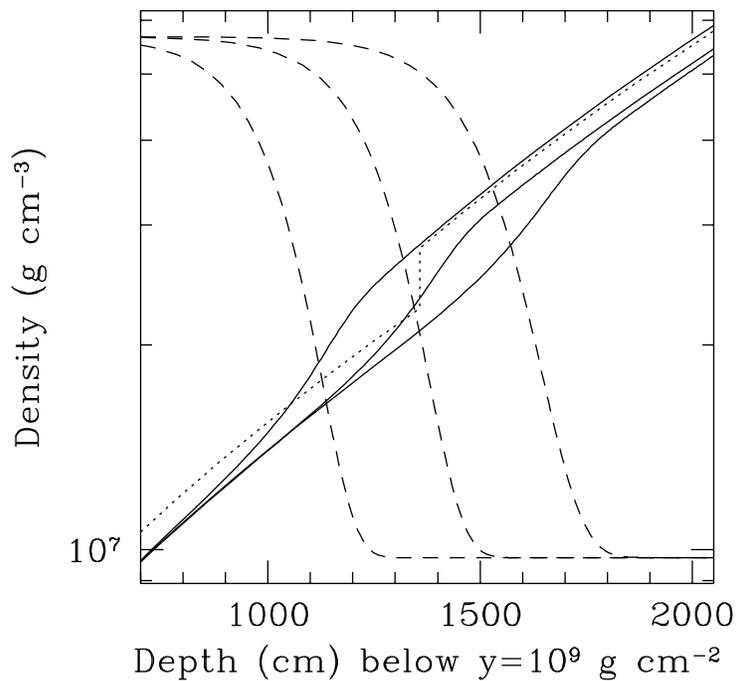,width=4 in}}
\caption{
Density as a function of depth in an isothermal ocean ($T_8=5$) with
an initial hydrogen mass fraction $X_r=0.1$. The solid lines show
(left to right) $\dot m/\dot m_{Edd}=0.01,0.1$ and $1$.  The dashed
lines (left to right) show the hydrogen mass fraction $X$ as a
function of depth on a linear scale, with the value at the left being
$X_r=0.1$. The dotted line is for a $T=0$ completely discontinuous
model where all captures occur at $E_F=Q$. The density
discontinuity mode for this density profile is shown in the left panel of
Figure \ref{fig:disc}.
\label{fig:rhoz}}
\end{figure}

The left panel in Figure \ref{fig:disc} displays the eigenfunctions
for a $T=0$ density discontinuity mode with $X_r=0.1$.  The density
profile used for this mode is the dotted line in Figure
\ref{fig:rhoz}. There is a 
large contrast (and actually a sign change) in the amplitude of the
transverse displacement above and below the discontinuity. This means
that almost all of the mode energy lies in the upper part of the
ocean, above the boundary layer. The radial displacement has maximum
amplitude at the discontinuity, and decays towards the ocean surface
and floor.

\subsubsection{Discontinuity Modes with a Finite-Thickness Transition Layer}
\label{sec:discfinite}

 We now consider models with a finite thickness hydrogen electron
capture layer that are appropriate for the accretion rates we are
considering. In these models, the density continuously changes across
the layer. The solid lines in Figure \ref{fig:rhoz} show the density
as a function of position for three different accretion rates when
$T_8=5$ and $X_r=0.1$. The dashed lines show the changing mass
fraction of hydrogen. The dotted line shows the case where all the
captures occur at $E_F=Q$ and $T=0$. Clearly, such a sharp
discontinuity is no longer present at these high accretion rates.
However, we find that the density discontinuity mode has the same
frequency as the discontinuous model as long as the electron captures
occur over less than a pressure scale height. The eigenfunction is
also qualitatively the same, except that the boundary layer is
resolved, for example there is a large transverse shear as
$\xi_\perp$ rapidly decreases\footnote{ One might worry that the large
transverse shear in the transition layer would be unstable to a
Kelvin-Helmholtz type instability. At the accretion rates and $X_r$'s
we consider, however, we find that the shear in the transition layer
is stabilized by the buoyancy there. The Richardson number in the layer
is always much greater than $1/4$ as long as the transverse
displacement at the top of the ocean is less than a stellar
radius.}, and the radial node which at $T=0$ lies
immediately below the discontinuity is moved deeper in the ocean.  The
right panel of Figure \ref{fig:disc} shows a typical discontinuity
mode in an isothermal ocean.

  It might come as a surprise that the same discontinuity mode appears
in both the simple model and the models with a finite thickness
electron capture layer. Indeed, one case involves matching conditions
across the discontinuity while the other involves smoothly integrating
the adiabatic mode equations through the layer. We now show that, in
the limit of a thin layer, the matching conditions can be recovered
by integrating the mode equations through the electron capture layer.
We use the transverse momentum equation, $ik\delta
p=\rho\omega^2\xi_\perp$ to transform the dependent variables in the
adiabatic mode equations (\ref{eq:mode1}) and (\ref{eq:mode2}) from
($\delta p/p,\xi_z$) to ($\xi_\perp,\xi_z$), finding
\begin{equation}\label{eq:tmode2}
{d\xi_z\over
dz}={\xi_z\over\Gamma_1h}+ik\xi_\perp\left({\omega^2\over c_s^2k^2}-1
\right)
\end{equation}
\begin{equation}\label{eq:tmode1}
{d\xi_\perp\over dz}={N^2\xi_\perp\over
g}+ik\xi_z\left(1-{N^2\over\omega^2}\right),
\end{equation}
where $c_s$ is the adiabatic sound speed, $c_s^2\equiv\Gamma_1gh$.
The discontinuity mode has $\omega_d^2\ll N^2$,
$\omega^2_d\ll c_s^2k^2$ and $\xi_z\lesssim kh\xi_\perp$, allowing
equation (\ref{eq:tmode2}) to be approximated as 
\begin{equation}
{d\xi_z\over
dz}={\xi_z\over
h}\left[{1\over\Gamma_1}-i(kh)\left({\xi_\perp\over\xi_z}\right)\right]
={\cal O}\left({\xi_z\over h}\right).
\end{equation}
Integrating this through the transition layer gives 
\begin{equation}\label{eq:match1}
\xi_{z-}=\xi_{z+}\left[1+{\cal O}\left({\Delta z\over h}\right)\right],
\end{equation}
where $\Delta z\ll h$ is the thickness of the layer. The pressure is
nearly constant within the electron capture layer as $\rho$ and
$\mu_e$ trade against each other. In other words the density scale
height is much less than the pressure scale height, $h(d\ln\rho/dz)\gg
1$, so we simplify 
\begin{equation}
N^2=-g\left(\dlnz{\rho}+{1\over\Gamma_1h}\right)\approx
-g\dlnz{\rho},
\end{equation}
and equation (\ref{eq:tmode1}) becomes
\begin{equation}
{d\xi_\perp\over dz}=
-\dlnz{\rho}\left(\xi_\perp-i{kg\over\omega^2}\xi_z\right).
\end{equation}
Using the fact that $\xi_z$ is nearly constant in the layer, we
integrate though the layer using the transverse momentum
equation to find 
\begin{equation}
{h_+\Delta p_+\over h_-\Delta p_-}={\rho_-\over\rho_+}.
\end{equation}
But since the pressure is constant, so is $\rho h$, giving us 
\begin{equation}\label{eq:match2}
\Delta p_-=\Delta p_+\left[1+{\cal O}\left({\Delta z\over h}\right)\right].
\end{equation}
For a thin boundary layer ($\Delta z\ll h$), equations
(\ref{eq:match1}) and (\ref{eq:match2}) are just the matching
conditions from our simple model. 

\subsubsection{Dependence of $f_d$ on Temperature and Accretion Rate}

The discontinuity mode frequency differs slightly from the simple
estimate above (equation [\ref{eq:fd}]) due to three different
effects: (1) pre-threshold electron captures change the position of
the layer in the ocean, (2) thermal buoyancy provides some extra
restoring force above and below the transition layer, and (3) at high
$\dot m$ and $T$ the thickness of the layer can exceed the local scale
height. However, we found that the dependence on these factors is
rather weak and typically is not larger than 20--30\% for an order of
magnitude change in $\dot m$ or $T$.
 
\begin{figure}[hbp]
\centering{\epsfig{file=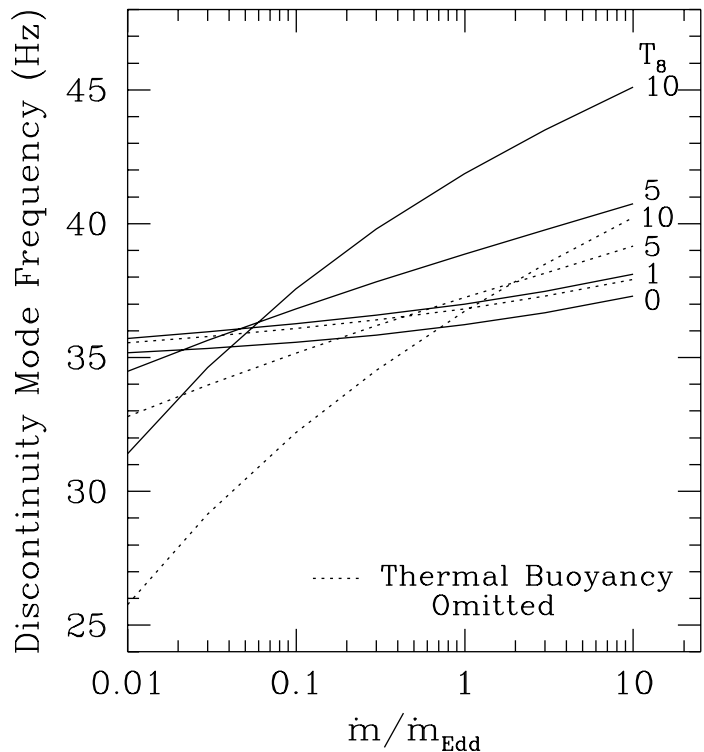,width=4 in}}
\caption{
The effect of accretion rate and temperature on the $l=1$
discontinuity mode frequency when $X_r=0.1$. The solid lines show
$f_d$ as a function of $\dot m$ for isothermal oceans with $T_8=0,1,5$
and $10$. The dotted lines show $f_d$ for models where the thermal
buoyancy is omitted.  At low $\dot m$, $f_d$ {\it decreases} with
increasing $T_8$.  This is because pre-threshold captures move the
electron capture layer upwards in the ocean, decreasing the scale
height at the discontinuity. At high $\dot m$, $f_d$ {\it increases}
with $T_8$ because of increasing thermal buoyancy and layer thickness.
\label{fig:fdm}}
\end{figure}

Figure \ref{fig:fdm} shows the effect of $\dot m$ and
$T$ on $f_d$ for isothermal oceans with $X_r=0.1$.
At low accretion rates ($\dot m\lesssim 0.05 \dot m_{\rm Edd}$), 
increasing $T_8$ {\it decreases} $f_d$ because pre-threshold 
captures move the electron capture layer upwards in the 
ocean, decreasing the scale height at the transition layer. 
At higher $\dot m$'s, $f_d$ {\it increases} with $T_8$ because of
increasing thermal buoyancy and thickness of the transition layer.
The effect of thermal buoyancy is shown by the dotted lines, which
show $f_d$ for models where the thermal buoyancy is omitted.
The effect of increasing layer thickness is shown by the $T=0$ line,
which can be fit by a simple $\dot m^{1/4}$ scaling, as suggested by 
equation (\ref{eq:physthick}), for the physical width of the electron capture
layer. In the relevant temperature range of $T_8\approx 5$--$7$, we find that
the frequency scales roughly as $f_d\propto \dot{m}^{1/36}$.

The models shown in Figure \ref{fig:fdm} are equilibrium 
models of the ocean. In reality, the electron capture layer is 
unlikely to be in complete thermal balance with the current 
value of the hydrogen mass fraction, as the thermal time 
there (minutes to hours) is much less than the accretion time (hours to days).
The response of the mode frequencies to changes in temperature on
a short timescale will be determined by changes in the thermal buoyancy and
not in the structure of the layer itself. 

\subsection{Trapped Modes within the Transition Layer}
\label{sec:trapmodes}

  We have also found a new set of g-modes which reside within the
electron capture layer. The restoring force for these modes comes
mainly from the buoyancy within the layer due to the composition
gradient, and so they exist even when $T=0$.  However, their existence
depends on the finite thickness of the electron capture layer, which
is a characteristic of the layers in accreting neutron stars. These
modes were not found in the density discontinuity mode studies of
isolated neutron stars (Finn 1987; McDermott 1990; Strohmayer 1993),
as in that environment the electron capture transitions were presumed
(most likely safely) to be sharp. Though the modes propagate above and
beyond the transition layer, we will refer to them as ``trapped'' modes,
as we can understand their frequencies depending on how many nodes are
present in the transition layer.

\begin{figure}[hbp]
\centering{\epsfig{file=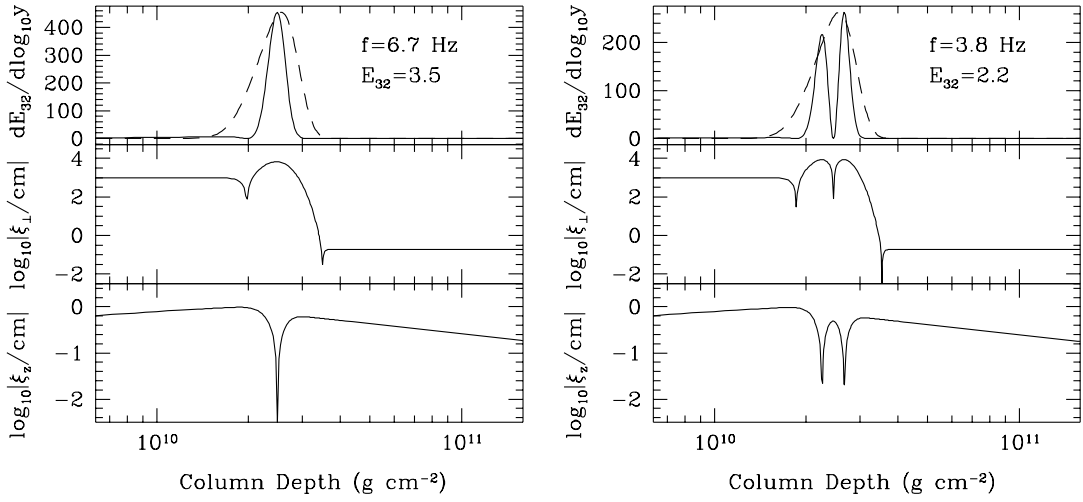,width=\hsize}}
\caption{
The $n_{tr}=1$ and $n_{tr}=2$, $l=1$, trapped modes in a $T=0$ ocean
with $\dot m=\dot m_{Edd}$ and $X_r=0.1$. The restoring force for this
mode is provided by the $\mu_e$ gradient in the electron capture
boundary layer. Almost all of the mode energy lies in the boundary
layer.  Labels and axes are the same as described in Figure
\ref{fig:disc}.
\label{fig:trap}}
\end{figure}

  Figure \ref{fig:trap} shows the $n_{tr}=1$ and $n_{tr}=2$ trapped modes in a
$T=0$ ocean, where $n_{tr}$ is the number of radial nodes {\it
within the transition layer}. Almost all of the mode energy lies
in the transition layer in the unphysical $T=0$ limit.  The transverse
displacement is constant in the upper and lower oceans because $N$
vanishes there and the mode is mostly transverse (see eq.
[\ref{eq:tmode1}]).  When there are many radial nodes per scale height
($k_zh\gg 1$) we can estimate the mode frequencies in the WKB
limit.  The coefficients in the adiabatic mode equations
(\ref{eq:mode1}) and (\ref{eq:mode2}) are then constant and are
integrated to obtain the dispersion relation for high order g-modes,
$k_z^2=N^2k^2/\omega^2$.  We require that there be an integer number
of wavelengths, $\int k_zdz=n\pi$, giving the
frequency of the $n$'th mode as
\begin{equation}\label{eq:WKB}
f_{n}\approx
{k\over 2\pi^2}{1\over n}\int Ndz=
{k\over 2\pi^2}{1\over n}\int Nh\ d\ln y.
\end{equation}
The trapped mode frequencies are set by the internal buoyancy due to
the composition gradient in the transition layer. For nearly all cases
we are considering, this dominates the thermal buoyancy within the
layer (see Figure \ref{fig:N2}), so that $N$ in that region is given
by
\begin{equation}
N^2=\frac{g}{h}\frac{d\ln\mu_e}{d\ln y}
=-\frac{g}{h}\mu_eX\frac{d\ln X}{d\ln y},
\end{equation}
where we have presumed (for simplicity) 
that the nuclear charge $Z$ is constant so that
$\mu_e$ is only a function of $X$. Adopting a relativistic equation of
state, we use the solution $X(E_F)=X(Q)\exp(-B)$, with $B$
given by equation (\ref{eq:B}), to evaluate $N$. Integrating through
the transition layer gives the convenient formula 
\begin{equation}\label{eq:ftr}
f_{tr}=\frac{8.5\Hz}{n_{tr}} 
\left({X_r\over 0.1}\right)^{1/2}
\left({\dot m\over \dot m_{\rm Edd}} {g\over 2\times 10^{14}\ {\rm cm\
s^{-2}}} {ft\over 1065 \ {\rm s}}\right)^{1/8}
\left({1293\ {\rm keV}\over Q}\right)^{5/8}
\times
\end{equation}
\begin{displaymath}
\left({10 \ {\rm km}\over R}\right)
\left({l(l+1)\over 2}\right)^{1/2},
\end{displaymath}
where we insert the prefactor found numerically. 
The analytic prefactor is close to that found numerically,
and is a function of fundamental constants.
We also show the scaling with $ft$ and $Q$,
since this formula is equally applicable to other electron capture boundary
layers. 

\begin{figure}[hbp]
\centering{\epsfig{file=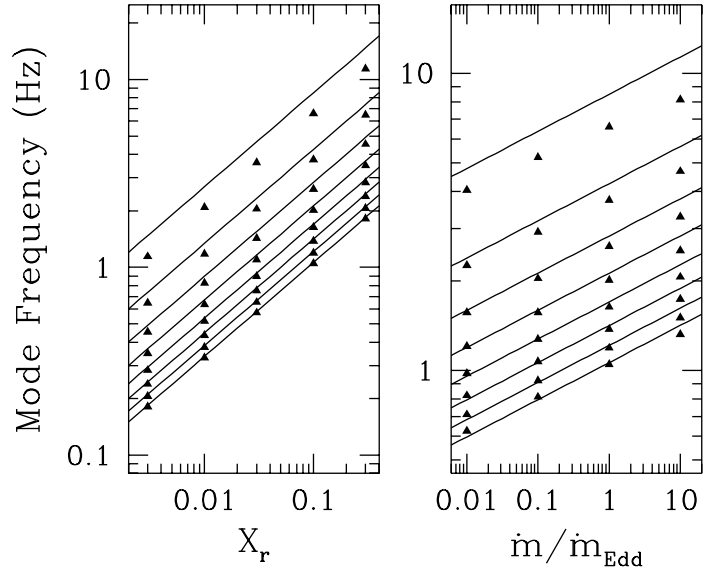,width=4 in}}
\caption{
The scalings of the $l=1$ trapped modes with $X_r$ and $\dot m$. The
left panel shows the frequencies of the first eight trapped modes
($n_{tr}=1$ has the highest frequency) in an ocean with $T=0$ and
$\dot m=\dot m_{Edd}$ ({\it triangles}) compared with the predicted
scaling $f_{tr}=8.5\ {\rm Hz}\ (X_r/0.1)^{1/2}/n_{tr}$ ({\it solid
lines}). The right panel compares the observed and predicted scaling
with $\dot m$, $f_{tr}=8.5\ {\rm Hz}\ (\dot m/\dot
m_{Edd})^{1/8}/n_{tr}$, when $X_r=0.1$ and $T=0$.  There is good
agreement, even for small $n_{tr}$ where the WKB approximation doesn't
strictly apply.  Here $n_{tr}$ is the number of radial nodes within
the transition layer.
\label{fig:ftz}}
\end{figure}

We find excellent agreement between the predicted scalings and our
numerical results. Figure \ref{fig:ftz} compares the analytic and
numerical scalings with $\dot m$ and $X_r$ for a range of $n_{tr}$
values. The left panel displays the first eight trapped mode
frequencies as a function of $X_r$ for a fixed accretion rate ($\dot
m=\dot m_{\rm Edd}$) and $T=0$. The filled triangles are our numerical
results and the solid lines are from equation (\ref{eq:ftr}).  The
right panel of Figure \ref{fig:ftz} compares the $\dot m$ scaling for
$X_r=0.1$ and $T=0$. It is clear that the WKB formulation gives an
excellent answer at high $n_{tr}$ and is reasonable at low $n_{tr}$ as
well.  We close by noting that the ratio of the discontinuity mode
frequency to a trapped mode frequencies is nearly independent of
everything
\begin{equation}
{f_d\over f_{tr}}=4n_{tr}
\left(1-\frac{\Delta Z}{\Delta A}\right)^{1/2}
\left({\dot{m}\over\dot{m}_{Edd}}\right)^{-1/8}
\left({g\over 2\times 10^{14}\ {\rm g\ cm^{-2}}}\right)^{-1/8},
\end{equation}
and thus might be an excellent observational discriminant.

\begin{figure}[hbp]
\centering{\epsfig{file=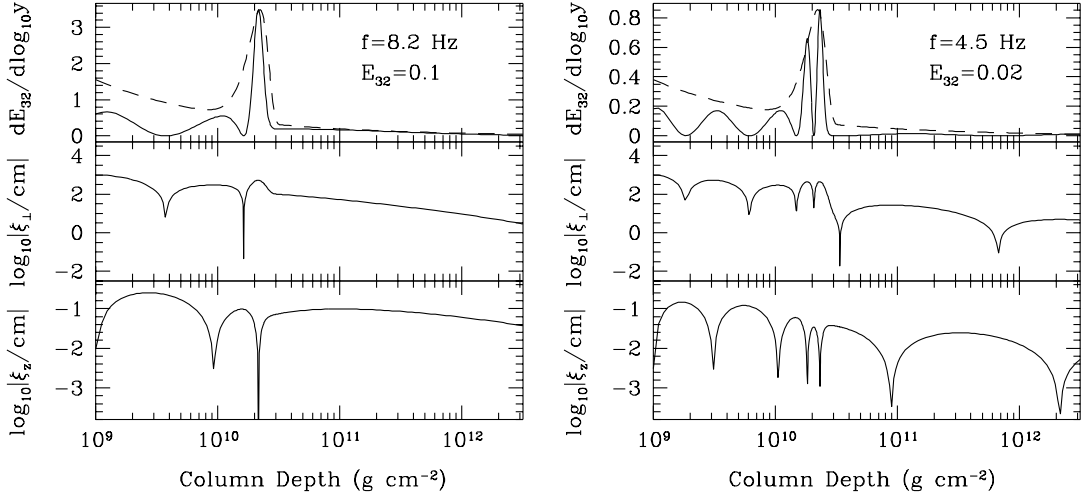,width=\hsize}}
\caption{
The $n_{tr}=1$ and $n_{tr}=2$, $l=1$ trapped modes in an ocean with $T_8=5$,
$\dot m=\dot m_{Edd}$ and $X_r=0.1$. The effect of thermal buoyancy is
to allow some of the mode energy to extend into the upper ocean.
Labels and axes are the same as described in Figure \ref{fig:disc}.
\label{fig:trap2}}
\end{figure}

For $T>0$, the effect of thermal buoyancy is to allow these modes
to extend into the upper ocean and atmosphere. Figure
\ref{fig:trap2} shows the $n_{tr}=1$ and $n_{tr}=2$ trapped modes in an
isothermal ocean with $T_8=5$. The frequencies are not much different
from their $T=0$ values.
The transverse displacement in the electron capture layer is less than
at $T=0$ because the thermal buoyancy above the layer makes
$\xi_\perp$ decay with depth.
The mode extension into the upper
ocean is critical for excitation of the trapped modes, particularly as
the mode then extends into the H/He burning layer. We are currently
investigating this possibility, as well as excitation from the
electron captures themselves.  The thermal buoyancy is also larger at
lower pressures and this will increase the mode frequency when the
mode amplitude is significant in the upper atmosphere. We now show
that at finite temperatures there is the additional complication due
to the coincidence that the thermal modes are ``mixed'' with these
trapped modes, at least in frequency space.

\subsection{Thermal Modes}
\label{sec:thmodes}

The thermal buoyancy throughout the deep ocean of degenerate electrons
produces a set of thermal g-modes (BC95). These modes were extensively
discussed by BC95, who pointed out the weak frequency dependence on the
depth of the ocean, a convenient situation since the depth depends so
strongly on uncertain quantities (see eq. [\ref{eq:crystal_y}]).
Just for completeness, we estimate the thermal g-mode frequencies in
the WKB limit by integrating equation (\ref{eq:WKB}). In a region
where $\mu_i$ is constant and the electrons are very degenerate,
equation (\ref{eq:N2therm}) gives $N^2 h^2\approx 3k_B T/8\mu_i
m_p={\rm constant}$ and the integration simply gives
\begin{equation}\label{eq:fth}
f_{th}={4.0\Hz\over n}
\left({T_8\over\mu_i}\right)^{1/2}
\ln\left(\frac{y_b}{y_t}\right)
\left({10 \ {\rm km}\over R}\right)
\left({l(l+1)\over 2}\right)^{1/2},
\end{equation}
where $y_t$ ($y_b$) is the column depth at the top (bottom) of the
ocean.  The frequencies depends most strongly on the temperature and ion
mean molecular weight and only logarithmically on the physical
thickness of the ocean.

\begin{figure}[hbp]
\centering{\epsfig{file=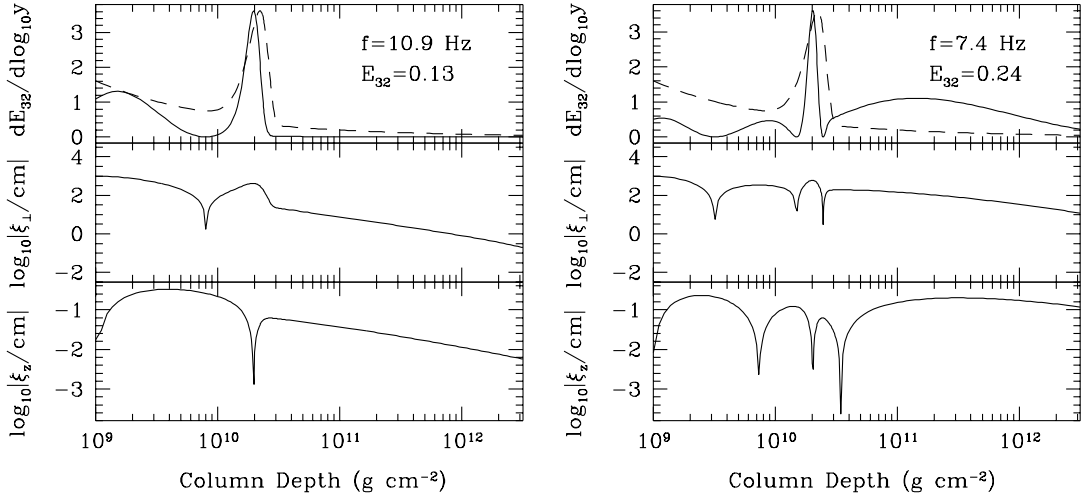,width=\hsize}}
\caption{Thermal g-mode $l=1$ eigenfunctions for an isothermal
ocean with $T_8=5$, $\dot m=\dot m_{\rm Edd}$ and $X_r=0.1$.  Left
(right) panel displays an upper (lower) thermal mode, so called as
most of the mode energy resides above (below) the electron capture
transition layer.  Labels and axes are the same as described in Figure
\ref{fig:disc}. 
\label{fig:therm}}
\end{figure}

 We find that the density discontinuity due to the hydrogen electron
captures divides the thermal g-modes into two types, which we
call the upper and lower thermal modes. Examples are shown in Figure
\ref{fig:therm} for an isothermal  ocean with $T_8=5$, $\dot m=\dot
m_{\rm Edd}$ and $X_r=0.1$. The upper thermal modes (left panel in
Figure \ref{fig:therm}) have most of their energy {\it above} the
electron capture layer. The layer acts as a floor for these modes and
there is a node in the radial displacement there, just as there would
be if we demanded our normal bottom boundary condition. This is not by
design, but is rather how the modes behave. This motivates us to use
our WKB estimate (eq. [\ref{eq:fth}]) and insert typical values
for $\mu_i$ and column depths in order to estimate these upper thermal
mode frequencies
\begin{equation}
f_{\rm th,upper}={9.0\Hz\over n_{\rm upper}}
\ 
\left({T_8\over 5}{9\over \mu_i}\right)^{1/2}
\ln\left[
\left({y_b\over 2\times 10^{10}\ {\rm g\ cm^{-2}}}\right)
\left({y_t\over 10^9\ {\rm g\ cm^{-2}}}\right)^{-1}
\right]\times
\end{equation}
\begin{displaymath}
\left({10 \ {\rm km}\over R}\right)
\left({l(l+1)\over 2}\right)^{1/2}.
\end{displaymath}
The lower thermal modes (example in the right panel of
Figure \ref{fig:therm}) have most of their energy {\it beneath} the
electron capture layer. 
We find that the Lagrangian pressure perturbation has a node at the
electron capture layer, just as it would if we put our traditional
top boundary condition at that location. In that sense, the
layer acts as a surface for these modes. Equation
(\ref{eq:fth}) then gives the lower thermal mode frequencies as
\begin{equation}
f_{\rm th,lower}={7.2\Hz\over n_{\rm lower}}
\
\left({T_8\over 5}{60\over \mu_i}\right)^{1/2}
\ln\left[
\left({y_b\over 10^{13}\ {\rm g\ cm^{-2}}}\right)
\left({y_t\over 2\times 10^{10}\ {\rm g\ cm^{-2}}}\right)^{-1}
\right]\times
\end{equation}
\begin{displaymath}
\left({10 {\rm km}\over R}\right)
\left({l(l+1)\over 2}\right)^{1/2},
\end{displaymath}
where we have put in the top column as the electron capture layer and
the bottom at the crystallization depth. We also took $\mu_i=60$, as
no hydrogen is present at these depths.  The upper and lower thermal
mode frequencies are similar because the difference in $\mu_i$ above
and below the electron capture layer is compensated by the greater
depth of the ocean beneath the electron layer than above. Here $n_{\rm
lower}$ ($n_{\rm upper}$) is the number of radial nodes below (above)
the electron capture transition layer.

The thermal mode spectrum is split because the jump in density and
$N^2$ across the transition layer creates an impedance mismatch, so
that an impinging wave is partially reflected by the transition
layer. This phenomenon is seen in so-called ``mode trapping'' in white
dwarfs, in which thermal modes are ``trapped'' in the thin hydrogen or
helium layers at the surface of the star (Winget, Van Horn \& Hansen
1981; Brassard et al. 1992). These modes (analogous to
our upper thermal modes) are believed to be preferentially excited.
It may well be that something similar happens in our case.
Just like the trapped modes (\S \ref{sec:trapmodes}), the 
upper thermal modes have significant
mode energy near the upper boundary, and so it is likely that they
will extend upwards into the upper atmosphere. We are currently
investigating this as we consider excitation mechanisms. 
The thermal buoyancy in the upper atmosphere will most likely increase the
frequency of the upper thermal modes. For example,
Strohmayer \& Lee (1996) investigated the seismology of an atmosphere
accreting and burning in steady state and found thermal g-modes of frequency
approximately $50 {\rm Hz}$ for comparable accretion rates. 
The lower thermal modes are more or less confined to the deeper
regions of the star. They do propagate upwards, but not enough to
have their frequencies modified by any atmospheric physics.

\subsection{Avoided Crossings and Mode Identification}\label{sec:avoid}

\begin{figure}[hbp]
\centering{\epsfig{file=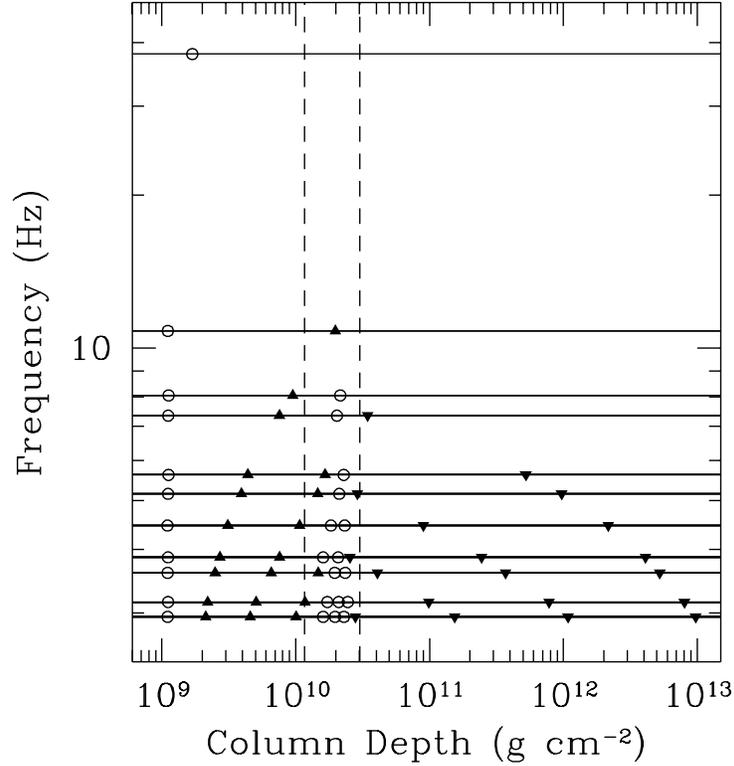,width=4 in}}
\caption{
The spectrum of $l=1$ modes in an isothermal ocean with $T_8=5$, $\dot
m=\dot m_{Edd}$ and $X_r=0.1$. Each horizontal line shows a mode
frequency. The highest frequency mode is the discontinuity mode, lower
frequency modes are trapped and thermal modes. The dashed vertical
lines show the places where the thermal and composition gradient
contributions to $N^2$ are equal; these lines show the approximate
boundaries of the electron capture layer. Moving down the diagram,
each new radial node is marked according to the type of mode, circles
for trapped modes and upward-pointing (downward-pointing) triangles
for upper (lower) thermal modes. The trapped nodes remain within the
boundary layer, whereas the nodes for the upper (lower) thermal modes
move upwards (downwards) in the ocean as the frequency decreases.
From the top down, the type of modes are d,u,t,l,u,l,t,l,u,t,l, where
d is for discontinuity, t is for trapped and u(l) is for an upper
(lower) thermal mode. See text for further discussion.
\label{fig:spec}}
\end{figure}

To conclude our discussion of isothermal models, we show in Figure
\ref{fig:spec} the spectrum of $l=1$ modes in an isothermal ocean with
$T_8=5$, $\dot m=\dot m_{Edd}$ and $X_r=0.1$, our fiducial isothermal
model.  Each horizontal line shows a mode frequency. The highest
frequency mode is the discontinuity mode, lower frequency modes are
trapped and thermal modes. The positions of the radial nodes in each
mode are shown as circles and triangles.  The dashed vertical lines
approximately bound the electron capture layer, showing where the
thermal and composition gradient contributions to $N^2$ are equal. 

Diagrams like the one in Figure \ref{fig:spec} help greatly in mode
classification.  The type of a given mode can be inferred more or less
by where the extra node is placed as one jumps from mode to mode down
the diagram.  Trapped modes have a new node added within the
transition layer and this node remains there as one moves down the
diagram. These nodes are marked with circles. A lower (upper) thermal
mode has a new node placed at the bottom (top) of the transition layer
and this node then moves downwards (upwards) in the ocean as one moves
down the diagram. These nodes are marked with downward-pointing
(upward-pointing) triangles. Thus in Figure \ref{fig:spec}, the types
of modes from top to bottom are d,u,t,l,u,l,t,l,u,t,l  where d is for
discontinuity, t is for trapped and u(l) is for an upper (lower)
thermal mode. This method of classifying nodes greatly eases the mode
identification problem. In particular, it helps to distinguish between
trapped and upper thermal modes, whose eigenfunctions may be very
similar\footnote{This is because the thermal buoyancy above the
transition layer begins to play a role in setting the trapped mode
frequencies. The trapped and upper thermal modes remain two distinct
sets of modes, however. For example, their frequencies depend
differently on the position of the upper boundary.} (for example,
compare Figures \ref{fig:trap2} and
\ref{fig:therm}). It is not completely reliable
however, as it is complicated by the mixing of mode eigenfunctions
near avoided crossings, as we now discuss.

It is by complete coincidence that the upper and lower thermal modes
are at a similar frequency to the trapped modes.  The frequencies of
these modes depend differently on $\dot m$, $T_8$, and $X_r$, leading
to ``avoided crossings'' whenever two mode frequencies try to cross
and an accidental degeneracy arises.  Avoided crossings were found for
non-radial oscillations in massive main sequence stars by Osaki
(1975), and in the neutron star context by Carroll et al. (1986), who
found avoided crossings between g-modes and p-modes in their
investigation of non-radial oscillations in the presence of strong
magnetic fields ($B\gtrsim 10^{11} \ {\rm G}$).

\begin{figure}[hbp]
\centering{\epsfig{file=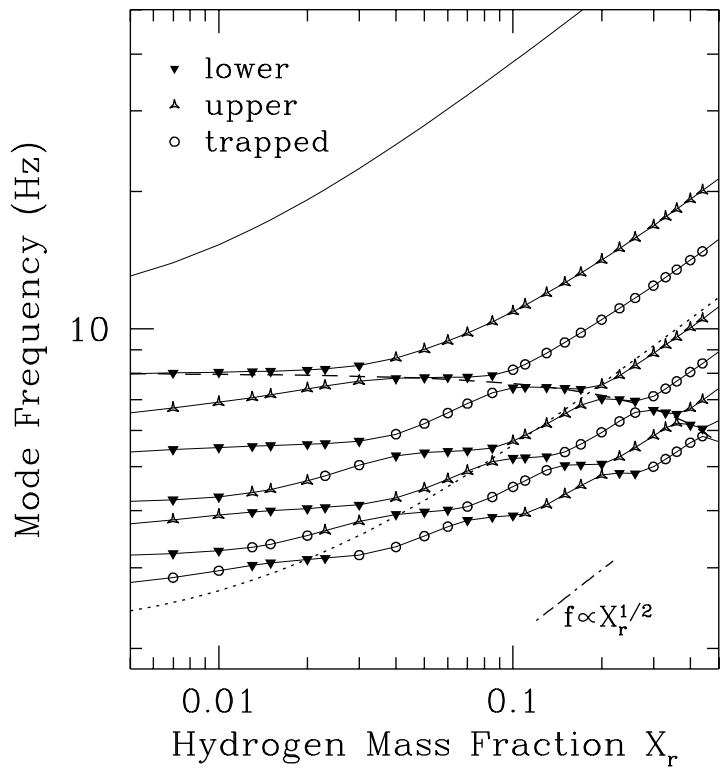,width=4 in}}
\caption{
The $l=1$ frequencies of the discontinuity mode and the first few
thermal/trapped modes as a function of hydrogen mass fraction $X_r$ in
an isothermal ocean with $T_8=5$ and $\dot{m}=\dot{m}_{Edd}$. Each
solid line is for a mode with a fixed number of radial nodes. The type
of mode is shown as a circle (trapped modes) or triangle (thermal
modes). The dot-dashed line is the $X_r^{1/2}$ scaling expected for
the trapped and discontinuity mode frequencies. The thermal mode
frequencies scale as $\mu_i^{-1/2}$. For the lower thermal modes
(dashed line) this gives a decreasing frequency with $X_r$, while for
the upper thermal modes (dotted line) the scaling is $\propto
X_r^{1/2}$ at high $X_r$ but levels out at low $X_r$.
\label{fig:fnvsX}}
\end{figure}

Avoided crossings can be seen in Figure \ref{fig:fnvsX} which shows
the frequencies of the discontinuity mode (upper solid line) and
several thermal and trapped modes as a function of $X_r$ for an
isothermal ocean with $T_8=5$ and $\dot m=\dot m_{\rm Edd}$.  Each
solid line is the frequency of a mode with a fixed number of radial
nodes $n$, and, as expected, the frequency always decreases as $n$
increases. The different dependencies of the mode frequencies on $X_r$
result in a series of avoided crossings. Along a solid line, the
number of radial nodes is conserved, but the character of the mode
(upper thermal, lower thermal or trapped, denoted by an
upward-pointing triangle, downward-pointing triangle or circle) may
change.  The dot-dashed line shows the $X_r^{1/2}$ scaling expected
for the trapped and discontinuity mode frequencies.  The thermal mode
frequencies scale as $\mu_i^{-1/2}$. For the lower thermal modes
(dashed line) this gives a decreasing frequency with $X_r$ because
$\mu_i$ in the lower ocean increases with $X_r$. For the upper thermal
modes (dotted line) the scaling is $\propto X_r^{1/2}$ at high $X_r$
but levels out at low $X_r$. On average, the different types of modes
follow the predicted scalings by changing $n$ in avoided crossings.
Locally, however, the behaviour with $X_r$ may be quite different.

\begin{figure}[hbp]
\centering{\epsfig{file=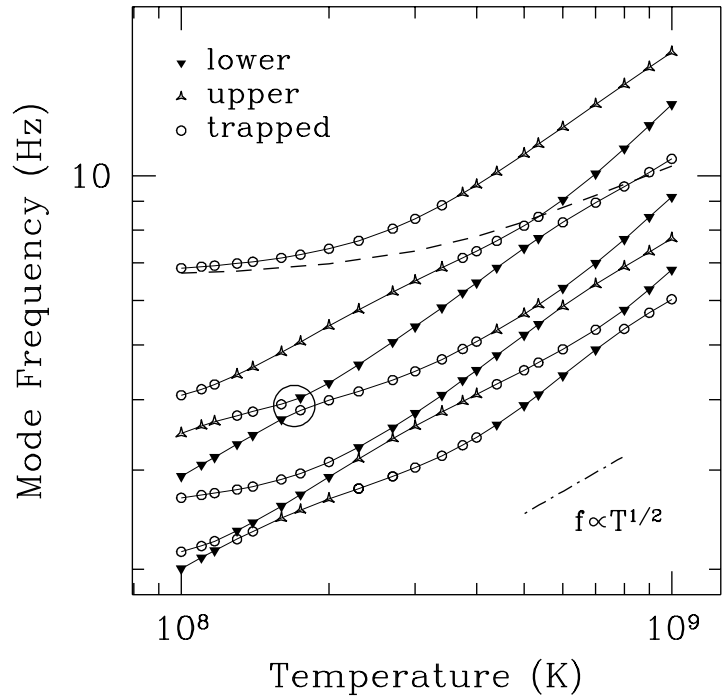,width=4 in}}
\caption{
The $l=1$ frequencies of the first few thermal/trapped modes as a
function of temperature in an isothermal ocean with $\dot m=\dot
m_{Edd}$ and $X_r=0.1$. As in Figure \ref{fig:fnvsX}, each solid line
is for a mode with a fixed number of radial nodes, and we show the
type of mode by a circle or triangle.  The dot-dashed line shows the
$T^{1/2}$ frequency scaling of the thermal modes. The dashed line
shows the frequency of the $n=1$ trapped mode when thermal buoyancy is
not included.  The circled avoided crossing is displayed in detail in
Figure \ref{fig:avoid}.
\label{fig:fnvsT}}
\end{figure}

Figure \ref{fig:fnvsT} is a similar diagram to Figure \ref{fig:fnvsX}
in which we show the dependence of the mode frequencies on temperature
in an isothermal ocean with $\dot m=\dot m_{Edd}$ and $X_r=0.1$.
Again, each solid line is for a mode with a fixed number of radial
nodes $n$, and, as before, the different scalings of the mode
frequencies with temperature yield avoided crossings. The dot-dashed
line shows the $T^{1/2}$ frequency scaling of the thermal modes. The
dashed line shows the frequency of the $n=1$ trapped mode when thermal
buoyancy is not included.  It shows how the frequency of the trapped
mode changes due to changes in the structure of the boundary
layer. Again, on average the modes follow the expected scaling with
$T$, but not necessarily locally.

Avoided crossings have been discussed by Aizenman et al. (1977),
Gabriel (1980) and Christensen-Dalsgaard (1981) using an approach
analogous to degenerate perturbation theory in quantum mechanics. In
this picture the frequency splitting at the point of closest approach
during a crossing is proportional to the ``overlap'' between the
eigenfunctions of the two modes (the off-diagonal matrix element in
quantum mechanics).  This may explain why an avoided crossing between
an upper thermal mode and a trapped mode has a larger frequency
splitting than an avoided crossing involving a lower thermal mode in
Figures \ref{fig:fnvsX} and \ref{fig:fnvsT}. The ``overlap'' between
upper thermal mode and trapped mode eigenfunctions is greater than
between an upper thermal or trapped mode and a lower thermal mode.

\begin{figure}[hbp]
\centering{\epsfig{file=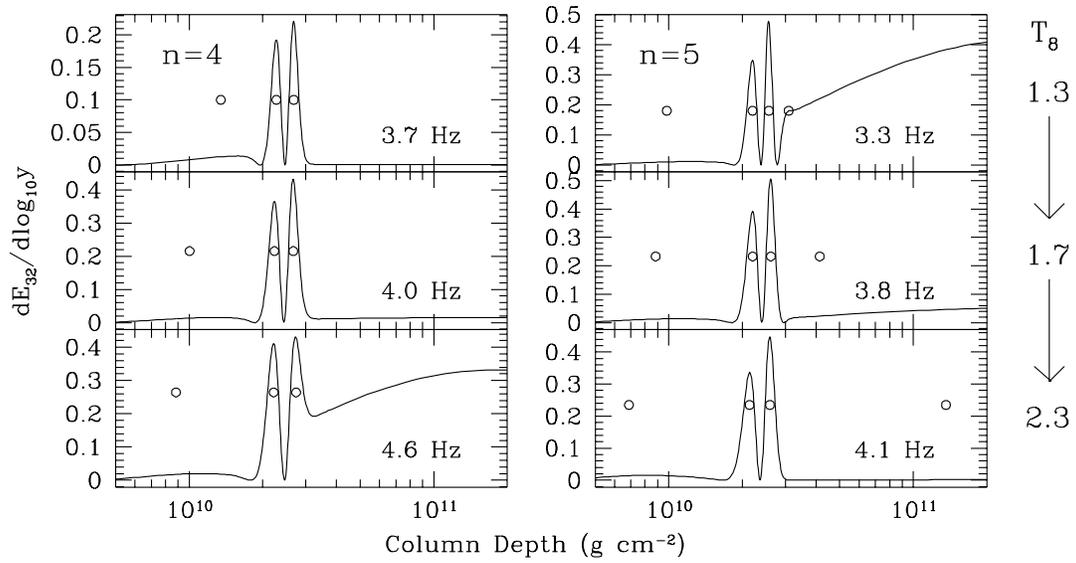,width=\hsize}}
\caption{
The avoided crossing (for $l=1$) between a trapped and lower thermal
mode. The $n=4$ and $n=5$ mode energy densities are shown as a
function of column depth for a range of temperatures from top to
bottom.  The circles show the position of the radial nodes. See text
for discussion. This avoided crossing is circled in Figure
\ref{fig:fnvsT}.
\label{fig:avoid}}
\end{figure}

During an avoided crossing, the mode eigenfunctions are mixed.  We
show this in Figure \ref{fig:avoid} in which we take the avoided
crossing which is circled in Figure \ref{fig:fnvsT}, and
show the energy density of the two modes before, during and after the
avoided crossing. This crossing is between a lower thermal mode and a
trapped mode. The $n=4$ and $n=5$ mode energy densities are shown as a
function of column depth for a range of temperatures from top to
bottom.  The circles show the position of the radial nodes. At
$T_8=1.3$, the lower frequency $n=5$ mode is a lower thermal mode, while
the higher frequency $n=4$ mode is a trapped mode. At the avoided
crossing, $T_8\approx 1.7$, the mode eigenfunctions are qualitatively
similar, differing only by a node. At higher temperature, $T_8=2.3$, the
two modes have exchanged characters. The lower frequency $n=5$ mode is
now a trapped mode while the higher frequency $n=4$ mode is a lower
thermal mode. The mixing of mode eigenfunctions may have implications
for mode excitation, allowing exchange of energy between the two
modes (for example, see Christensen-Dalsgaard 1981).

\section{Non-Isothermal Ocean Models} 
\label{sec:models}

Up to this point we have presumed that the ocean is isothermal and
considered the temperature to be a free parameter. We now construct
non-isothermal models of the ocean by integrating the heat equation
(see Brown \& Bildsten 1998) through the ocean.  There are several
sources of energy that contribute to the heat flux in the ocean.  The
first is the energy released when the neutrons from the hydrogen
electron captures combine with the heavy nuclei, roughly $E_H\approx
7\times 10^{18} \ {\rm ergs \ g^{-1}}\approx 7$ MeV per accreted
nucleon. The second is the energy released as matter falls downwards
(or what is sometimes called ``compressional heating''), $E_{\rm
comp}\sim k_BT/\mu_i m_p$. Finally, there is energy released in the
deep crust by nuclear electron captures and pycnonuclear
reactions. Most of this energy (about an MeV per accreted nucleon,
Haensel \& Zdunik, 1990a) goes into the NS core and is lost via
neutrinos, about 10\% ($E_{cr}\sim 0.1$ MeV per accreted nucleon)
flows upwards through the ocean (Brown \& Bildsten 1998). The total
heat flux above the electron captures (at say $y\approx 10^9
\ {\rm g \ cm^{-2}}$) is
\begin{equation}\label{eq:fluxcomp}
F_{\rm deep}\approx \dot m(E_H X_r+ E_{\rm comp}+E_{cr}).
\end{equation}
The largest uncertainty in $F_{\rm deep}$ is the amount of hydrogen left
unburned and advected downward, $X_r$ (Taam et al. 1996; Schatz et
al. 1997b).  As we show in this section (in agreement with Taam et
al. 1996), for typical values of $X_r\sim 0.1$ and high accretion
rates the region between the H/He burning layer and the hydrogen
electron captures has a substantial temperature gradient. The other
contributions to $F_{\rm deep}$ are typically not large enough to require
a substantial temperature gradient (BC95) at sub-Eddington accretion
rates, especially in the relativistic ocean underneath the hydrogen
electron captures.

Even when $F_{\rm deep}$ is known and nuclear reactions beyond electron
capture are ignored\footnote{For the purposes of illustration in this
section, we persist with our simple model of electron capture followed
by neutron capture onto a single type of nucleus. There is no reason
to undertake a more complete treatment of the neutron
captures/$\beta$-decays and the possibility of direct proton captures
onto the nuclei when $T>10^9 \ {\rm K}$ until we have better knowledge
of the nuclear mix from the H/He burning.}, there are still additional
uncertainties. One is the radiative opacity for this complicated
mixture of elements. For pure iron at the depth where the hydrogen
electron captures are occurring, the predominant heat transport
mechanism would be electron conduction (Gudmundsson, Pethick, \&
Epstein 1983). Thus for simplicity, we neglect radiative heat
transport and assume that all the heat is carried by electron
conduction.  For the conductive opacities, we use the results of
Yakovlev \& Urpin (1980) for electron-ion collisions, and the fit of
Potekhin et al. (1997) for electron-electron collisions.  An
additional uncertainty is the choice of temperature at the top of the
ocean when the H/He burning is time dependent. This is because the
separation between the electron capture region and the burning
location is not far enough that a simple radiative-zero like solution
applies, at least when $T\gtrsim 5\times 10^8 {\rm K}$ at the electron
capture depth.

\subsection{A Few Illustrative Models}
\label{sec:nonthmod}

\begin{figure}[hbp]
\centering{\epsfig{file=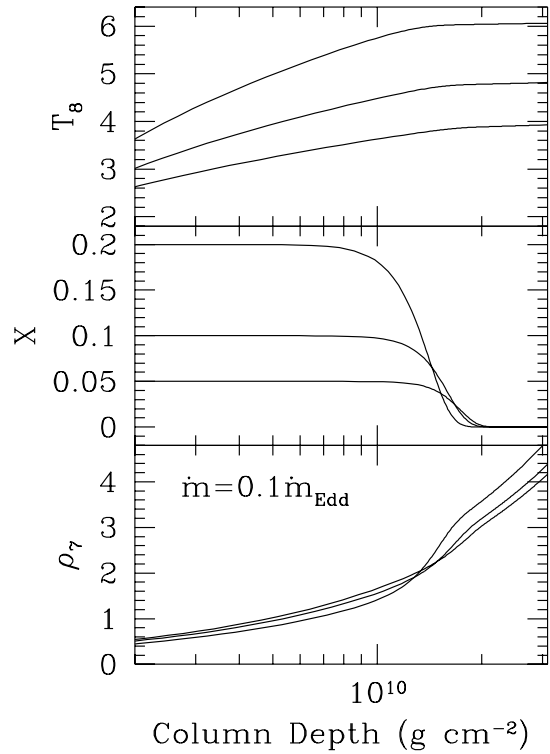,width=3 in}}
\caption{
The structure of an $R=10 \ {\rm km}$, $M=1.4 M_\odot$ NS near the
hydrogen electron capture region for an
accretion rate of $\dot m=0.1 \dot m_{\rm Edd}$ and different values
of $X_r$. The three lines in the top (middle) panel show the
temperature in units of $10^8 \ {\rm K}$ (hydrogen mass fraction, $X$)
as a function of the column depth $y=p/g$ for (from top to bottom)
$X_r=0.2,0.1$ and 0.05. The bottom panel displays the density (in
units of $10^7 \ {\rm g \ cm^{-3}}$) for the three different values of
$X_r$. The line with the highest density at the highest column is for
$X_r=0.2$.
\label{fig:struct1}}
\end{figure}

In order to illustrate the effects of $F_{\rm deep}$, we show in Figure
\ref{fig:struct1} the temperature, hydrogen mass fraction and
density as a function of column depth for models with $\dot m=0.1 \dot
m_{\rm Edd}$ and $X_r=0.2, 0.1$ and $0.05$ (from top to bottom in the
$T$ and $X$ panels). In the density panel, the curve with the highest
density at the largest column depth has $X_r=0.2$. These models have
the outer boundary condition $T=2\times 10^8 \ {\rm K}$ at $y=10^9 \
{\rm g \ cm^{-2}}$ and, for simplicity, we set $E_{\rm comp}=0$ and
$E_{cr}=10^{17} {\rm erg \ g^{-1}} \approx 0.1$ MeV per accreted
nucleon. The initial nucleus has $A_i=60, Z_i=30$ and is allowed to
become arbitrarily neutron rich. The flux at depths far below the
electron capture region is $F_{\rm bottom}=\dot m E_{cr}= 7.5 \times
10^{20}\ {\rm erg \ s^{-1} \ cm^{-2}}$. It is clear that, for these
values of $X_r$, the ocean is substantially hotter than if there were
no hydrogen present. At this accretion rate, the ocean is nearly
isothermal below the hydrogen electron captures.  Most of the electron
captures are post-threshold ($E_F>Q$) and occur 15--30 days after the
matter has arrived on the star.

\begin{figure}[hbp]
\centering{\epsfig{file=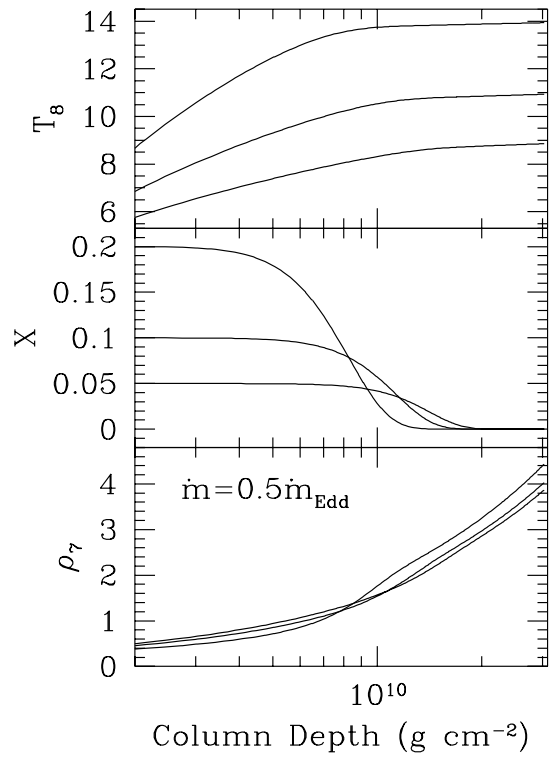,width=3 in}}
\caption{
The NS structure near the hydrogen electron 
capture region for $\dot m=0.5 \dot m_{\rm Edd}$ and different
values of $X_r$. The labels and arrangement are the same as in Figure
\ref{fig:struct1}.
\label{fig:struct2}}
\end{figure}

Figure \ref{fig:struct2} shows models with a higher accretion rate,
$\dot m=0.5 \dot m_{\rm Edd}$, and an outer boundary condition of
$T=4\times 10^8\ {\rm K}$ at $y=10^9 \ {\rm g \ cm^{-2}}$. The other
variables are the same as in the previous example. The enhanced flux
due to the higher accretion rate (see eq. [\ref{eq:fluxcomp}]) makes
these models much hotter than the lower accretion rate ones.  We may
have slightly overestimated the temperature at these high $\dot m$'s,
since for $T\gtrsim 10^9 \ {\rm K}$, radiation may start to carry a
significant fraction of the flux. Many of the electron captures for
the $X_r=0.2$ and $X_r=0.1$ cases occur pre-threshold ($E_F<Q$). For
the $X_r=0.2$ case, one half of the hydrogen is depleted only 2.5 days
after arriving on the NS.

At these high temperatures ($> 10^9 \ {\rm K}$) it is important to
consider neutrino cooling. Using the formulae given by Schinder et
al. (1987), we find that, even for the hottest model we consider here
($\dot m=0.5$, $X_r=0.2$), neutrino cooling is unimportant at the
depth of the hydrogen electron captures.  It is certainly very
important at greater depths however (Brown \& Bildsten 1998). Once
$y\sim E_{cr} \dot m/
\epsilon_\nu$, where $\epsilon_\nu$ is the neutrino cooling rate in
${\rm erg\ g^{-1}\ s^{-1}}$, the temperature gradient changes direction
and heat starts to flow into the core, a possibility we neglect for
now. For $\dot m=0.5$, this depth is $y\sim 10^{13}$--$10^{14} \ {\rm g \
cm^{-2}}$, near the bottom of the ocean. At higher $\dot m$'s or
$X_r$'s neutrino cooling will become important at even lower column
depths than this, although we stress again that at higher
temperatures it becomes important to include the radiative opacity.

How does the hydrogen electron capture transition layer compare
to the isothermal cases we calculated previously? Figures
\ref{fig:struct1} and \ref{fig:struct2} show that the transition layer
itself is nearly isothermal. For example, even in the model with $\dot
m=0.5$ and $X_r=0.2$, the temperature changes by a factor of only 10\%
across the transition layer.  We thus characterise the structure of
the layer by the temperature at, say, the place where one half of the
hydrogen is depleted. For example, the hydrogen mass fraction $X$ in
the non-isothermal model with $\dot m=0.5$ and $X_r=0.1$ agrees to
within 10\% at each depth with an isothermal model with $T=10^9\ {\rm
K}$ and the same $\dot m$ and $X_r$.

These examples are intended to illustrate the large impact of the
hydrogen electron captures (Taam et al. 1996) on the deep structure of
the neutron star. Much more work is needed on the products of H/He
burning before any definitive results can be stated. However, we hope
we have strengthened the case of Taam et al. (1996) that even small
amounts of residual hydrogen at high $\dot m$'s can play an important
role. We now describe the g-mode spectra of these non-isothermal
models.

\subsection{g-Mode Spectra for Non-Isothermal Models}

\begin{figure}[hbp]
\centering{\epsfig{file=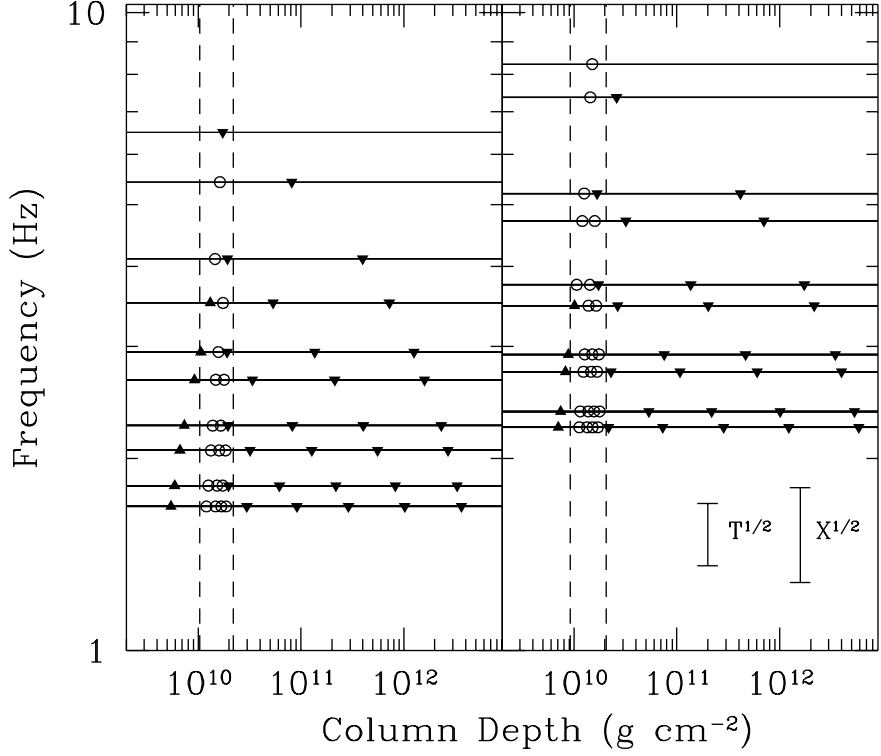,width=6 in}}
\caption{
The spectrum of $l=1$ modes in non-isothermal oceans with $\dot
m=0.1\dot m_{Edd}$ and $X_r=0.05$ (left panel) and $X_r=0.1$ (right
panel). The temperature at the top ($y=10^9\ {\rm g\ cm^{-2}}$) is
$T_8=2$ and rises to $T_8\approx 4$ ($X_r=0.05$) and $T_8\approx 5$
($X_r=0.1$) below the electron capture layer (at $y=3\times 10^{10}\
{\rm g\ cm^{-2}}$).  We have included a deep flux of 0.1 MeV per
accreted nucleon.  The initial nucleus before electron capture has
$A_i=60$, $Z_i=30$.  In each case, we show the discontinuity mode and
the first ten trapped and thermal modes. The symbols in this Figure
are the same as in Figure \ref{fig:spec}. For clarity, we have omitted
the first radial node which lies near the upper boundary. The vertical
bars show the difference in frequency expected between the left and right panels given the simple scalings $X_r^{1/2}$ and $T^{1/2}$.
\label{fig:doubleprop}}
\end{figure}

The g-mode spectra of non-isothermal models of the ocean are easy to
understand in the context of our study of isothermal ocean g-modes in \S
\ref{sec:modes}. The g-mode spectra for the $\dot m=0.1, X_r=0.05$ and
$\dot m=0.1, X_r=0.1$ models discussed above are shown in Figure
\ref{fig:doubleprop}.
The discontinuity mode frequency is insensitive to details of the
structure of the transition layer and depends mainly on $X_r$. It is
given accurately by equation (\ref{eq:fd}).  The lower thermal mode
frequencies are set by the thermal buoyancy in the ocean below the
transition layer, which is almost isothermal. The frequencies are
almost the same as the lower thermal mode frequencies in an isothermal
ocean with a temperature equal to that just below the transition
layer.  The discontinuity and lower thermal mode frequencies are
insensitive to details of the upper boundary and thermal buoyancy
above the transition layer. This is not true of the upper thermal and
(to a lesser extent) the trapped modes. Hence we expect that their
frequencies will change as we extend our models into the upper
atmosphere. Also, we have not included radiative opacity which becomes
important near the upper boundary and will reduce the temperature
gradient there. In Figure \ref{fig:doubleprop} we show the predicted
frequency increases due to temperature for the thermal modes
($T^{1/2}$ scaling) and due to $X_r$ for the trapped modes ($X^{1/2}$)
by vertical bars. These simple scalings predict the difference in
frequencies between the left and right panels of Figure
\ref{fig:doubleprop} fairly well.

\begin{figure}[hbp]
\centering{\epsfig{file=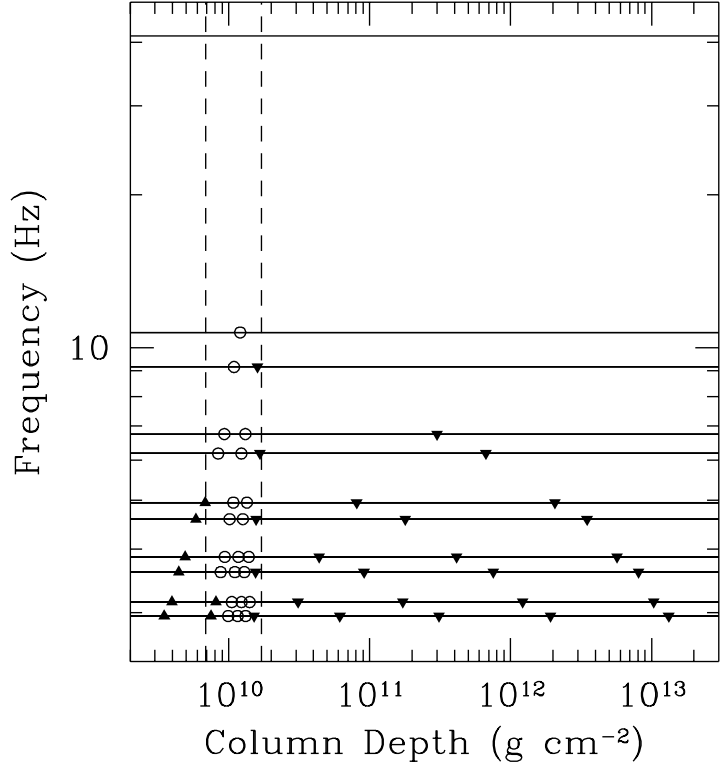,width=4 in}}
\caption{
The spectrum of $l=1$ modes in a non-isothermal ocean with $X_r=0.137$
and $\dot m=0.1\dot m_{Edd}$. The temperature at the top ($y=10^9\
{\rm g\ cm^{-2}}$) is $T_8=3.5$ and rises to $T_8=5.6$ below the electron
capture layer (at $y=3\times 10^{10}\ {\rm g\ cm^{-2}}$). 
We have included a deep flux of 0.1 MeV per accreted nucleon.
The initial nucleus before electron capture is iron ($A_i=56$, $Z_i=26$).
This model is based on model 8 of Taam et
al. (1996). We show the discontinuity mode and the first ten trapped
and thermal modes. The symbols in this Figure are the same as in
Figure \ref{fig:spec}. For clarity, we have omitted the first radial
node which lies near the upper boundary.
From top to bottom, the modes are d,t,l,t,l,u,l,t,l,u,l, where d is for
discontinuity, t is for trapped and u(l) is for an upper (lower)
thermal mode. 
\label{fig:taam}}
\end{figure}

As a further example, we have reconstructed one of the models of Taam
et al. (1996). They followed the time dependent evolution of models
with accretion rates ranging from $0.1$--$1$ times the Eddington rate.
All of these models had substantial residual hydrogen, ranging at the
highest $\dot m$'s from $X_r=0.08$--$0.21$.  Their neutron star had
$M=1.4 M_\odot$ and $R=9.1 \ {\rm km}$ and the initial heavy nucleus
was iron ($A_i=56, Z_i=26$).  We chose model 8 from their tables 1 \&
2 which has $X_r=0.137$ and $\dot m=0.1\dot m_{Edd}$, and varied our
outer boundary condition until we matched their stated deep
temperature.  We then calculated the g-mode spectrum of this model.
This is shown in Figure \ref{fig:taam}.

\section{Conclusions}
  
We have exhaustively studied how the hydrogen electron capture layer
expected in the rapidly accreting ($\dot M> 10^{-10} M_\odot \ {\rm
yr^{-1}}$) neutron stars affects the internal g-mode spectrum of these
objects.  The abrupt rise in density across the layer supports a
density discontinuity mode (see \S \ref{sec:dismodes}) of frequency
\begin{equation}\label{eq:fdintro}
f_d\approx 35\ {\rm Hz}
\left({X_r}\over {0.1}\right)^{1/2}\left(1-{\Delta Z\over \Delta
A}\right)^{1/2} 
\left({10 \ {\rm km}}\over R\right)
\left( {l(l+1)\over 2} \right)^{1/2}, 
\end{equation}
where $\Delta Z$ and $\Delta A$ are the average change in the nuclei's
charge and mass through the layer due to neutron captures and
subsequent $\beta$-decays.  The internal buoyancy due to the
composition gradient within the electron capture boundary layer
creates a new spectrum of ``trapped'' modes (see \S
\ref{sec:trapmodes}) of frequency
\begin{equation}\label{eq:fnintro}
f_{tr}\approx \frac{8.5 \ {\rm Hz}}{n_{\rm tr}}
\left({X_r}\over {0.1}\right)^{1/2}
\left({10 \ {\rm km}}\over R\right)
\left( {l(l+1)\over 2}\right)^{1/2},
\end{equation}
where $n_{\rm tr}$ is the number of nodes in the boundary layer and we have
omitted the weak dependence on the accretion rate. 
There is also a set of thermal g-modes in the same frequency range as
the trapped modes, which are separated by the density
discontinuity into two distinct sets (see \S \ref{sec:thmodes}).

This work might eventually provide a natural explanation for some of
the observed QPO's in accreting NS. Indeed, if the neutron stars in
the ``Z'' sources were slowly rotating, the discontinuity (trapped)
modes from the hydrogen electron capture layer would nicely overlap
the freqencies of the QPOs seen in the horizontal (normal) branch. The
stable 40 Hz QPO in the $f_s\approx 2\ {\rm Hz}$ X-ray pulsar GRO
J1744-28 (Zhang et al. 1996) is an intriguing one to consider, as the
discontinuity mode is in the right frequency range and might have a
small enough shear so that the $\lesssim 10^{11} \ {\rm G}$ field of
this pulsar (Bildsten \& Brown 1997) will not affect the mode
frequency. However, there is still much theoretical work to be done,
from understanding how the modes are excited to how they modulate the
X-ray flux.

\subsection{Observational Tests}

Observational progress will come by identifying QPOs with modes of
different $l$'s, where the ratio of the mode frequencies are known.
This is easiest when the neutron star is rotating slowly (spin
frequency $f_s < 10$ Hz), as one then merely looks for the $\omega^2
\propto l(l+1)$ scaling between QPO's that are (ideally)
simultaneously present. The rapidly rotating case is more complicated
as the angular eigenfunctions and dispersion relations are very
different. As we discuss below, seismological progress can most likely
be made only for those NS where the spin frequency has been measured
during a Type I burst (see Bildsten 1998b for a summary of those
objects).

An important question is how fast can the mode frequencies change? The
mode frequencies depend on internal conditions in the ocean, in
particular on the temperature and $X_r$. If a QPO was observed to
change its frequency faster than the internal conditions in the ocean
can change or if the dynamic range of the frequency was too great,
this could rule out non-radial oscillations as a source of the
periodicity.
The discontinuity mode frequency and the trapped mode frequencies are
sensitive to the residual hydrogen mass fraction, $X_r$. This changes
from one X-ray burst to another, which will result in layers of
different $X_r$ being compressed towards the electron capture boundary
layer. These layers are most likely Rayleigh-Taylor unstable and will
mix, making the timescale for a large frequency change roughly the
accretion time at the transition layer, days to weeks.  The thermal
modes and trapped modes are sensitive to the temperature in the
ocean. The important timescale is the thermal time at the place where
most of the mode energy resides, hours to days at the electron capture
depth. Thus the upper thermal mode frequencies will change faster than
the lower thermal mode frequencies. As we noted in \S \ref{sec:modes},
since the thermal time is shorter than the accretion time, the change
in the mode frequencies on a short timescale will be determined by the
change in the thermal buoyancy and not the structure of the transition
layer.

\subsection{The Rapidly Rotating Case}

There is good reason to suspect that these neutron stars are rapidly
rotating, as the prolonged accumulation of material at this rate will
most likely spin up the star. The coherent periodicities during Type I
X-ray bursts seem to indicate 250--500 Hz spin frequencies, which are
much greater than the g-mode frequencies, but still small compared to
the breakup frequency $\Omega_b\approx (GM/R^3)^{1/2}$ ($\approx 2$
kHz for a $1.4 M_\odot, R=10$ km star). When $\Omega \ll \Omega_b$,
the unperturbed star is spherical and the centrifugal force can be
neglected, in which case the primary difference in the momentum
equations is the Coriolis force. As a result, the g-mode frequencies
depart significantly from the $\omega^2 \propto l(l+1)$ scaling
(Papaloizou \& Pringle 1978).
 
BUC96 made progress on this problem within what is called the
``traditional approximation'', where the radial and transverse
momentum equations separate and the resulting angular equation must be
solved to find the angular eigenfunctions (no longer $Y_{lm}$'s) and
transverse eigenvalues $\lambda\equiv kR^2$ (no longer $l(l+1)$).
The radial equations are identical to the non-rotating case, so that
if $\omega_0$ is the eigenfrequency for the $l=1$ mode of a {\it
non-rotating} star, then the oscillation frequency (in the rotating
frame) at arbitrary spin is $\omega=\omega_0(\lambda/2)^{1/2}$.  These
are then transferred into the observer's inertial frame via
$\omega_I=\omega-m\Omega$, so that going from our non-rotating results
to the rotating star is straightforward. However, nearly any frequency
can be predicted in the absence of prior knowledge of the spin
frequency, as the retrograde modes cover a large frequency range as
the NS spin is varied.

Robust predictions are, however, possible for those NSs where the spin
frequency is known from a Type I burst, By far the best case to date
is 4U~1728-34 (Strohmayer et al. 1996; Strohmayer, Zhang \& Swank
1997) which has $f_s=363$ Hz. For a known spin frequency, BUC96 and
Bildsten et. al. (1998) have shown that one can predict the observed
frequencies as a function of the internal conditions (e.g. $T$ or
$X_r$). Bildsten et al. (1998) asked how the observed frequencies for
different angular eigenmodes would change as the internal conditions
change. They found that the prograde modes are all at relatively high
frequencies, near a kHz. However, these mode frequencies did not
exhibit the large dynamic range of the observed kHz QPOs. The
splitting for some of the prograde modes was found to be nearly
constant, but {\it not} equal to the spin frequency. Other prograde
modes had varying differences and could potentially be applicable to
those kHz QPO's that do not have constant frequency separation (Sco
X-1, van der Klis et al. 1997, 4U~1608-52, Mendez et al. 1998). Of
particular interest were the retrograde modes that appeared at $f<100
\ {\rm Hz}$. These showed a large dynamic range for only a small
change in internal conditions. Bildsten et al. (1998) thus pointed to
the possibility of explaining some of the QPOs seen in the $<100$ Hz
range from 4U~1728-34 and other Atoll sources.

Bildsten et al. (1998) pointed out that confirming non-radial
pulsations in a star with a previously measured spin frequency can
come by identifying a measured low-frequency QPO with a particular
value of the non-rotating frequency and then searching in the data for
the higher frequency prograde modes. Such an exercise has yet to be
carried out and could potentially bear fruit, as few Fourier bins
would need to be searched. The dispersion relations are complicated
enough that such exercises can only be done in concert with theory. At
present, theoretical predictions can most likely be done at the 10\%
level. Further theoretical work to include general relativistic
effects might be needed for the very rapidly rotating stars ($f_s> 500
\ {\rm Hz}$) before such an exercise could be carried out in detail,
say at the 1\% level.

\subsection{Future Work on the Capture Layer, Mode Excitation and Coupling to Accretion} 

There is still much work needed on the physics of the hydrogen
electron capture layer. For example, a more realistic calculation
would follow several species of heavy nuclei and incorporate neutron
capture cross sections and $\beta$ decay half-lives for each. It is
important to carry out some of these calculations, as there is a
possible large-scale instability that might develop underneath the
layer. After all the electron captures have occurred, the nuclei will,
on average, be neutron rich. These nuclei advect downwards and
eventually $\beta$-decay (they typically are not Fermi blocked) on
timescales much longer than days. This may cause a density inversion as the
released electrons will decrease the density, possibly leading to a
Rayleigh-Taylor instability underneath the hydrogen electron capture
zone.

 Which modes does the star like to excite? We are presently working on
the internal excitation and damping of the adiabatic modes we have
presented here. Following the work of SL96, we are starting by
considering the excitation of the upper thermal modes, trapped modes
and discontinuity modes by the H/He burning layers. Our new
understanding of the mode structure will modify the competition
between excitation in the burning layers and damping elsewhere in the
star that they found. We hope to show that a sub-set of the infinite
spectrum of oscillations we have found here will be excited. Finding
the steady-state amplitude of the oscillation is beyond the scope of
these calculations since it would require going beyond linear order.

 How can a mode modulate the X-ray luminosity? There are a few ways
one can imagine. One pointed out by Bildsten et al. (1998) for the
rapidly rotating case is from direct interaction with the accretion
disk, as the g-modes are confined to a narrow equatorial band that
would lie in the same plane as the accretion disk. If the NS does not
lie inside the last stable orbit, then the disk should run directly
into the surface, allowing for potential periodic modulation of the
accretion environment. An alternative is to invoke a weak magnetic
field. The mode frequencies are not modified when $B<10^9 \ {\rm G}$
(BC95), but the thermal modes have tremendous shear and will thus, at
high altitudes, change their nature (Carroll et al. 1986). Whether the
resulting outgoing waves can affect the accretion flow remains an open
question. 

We thank Hendrik Schatz and Michael Wiescher for conversations
regarding the nuclear physics, A. Muslimov for suggesting we check for
shear instability, and Greg Ushomirsky and Ed Brown for many
insightful conversations and comments on the manuscript. This work was
supported by the NASA Astrophysics Theory Program through grant NAG
5-2819.  L. B. was also supported by the Alfred P. Sloan Foundation.

\newpage

\newpage

\newpage

\end{document}